\documentclass[10pt,a4paper,twocolumn]{article}

\usepackage{authblk}
\usepackage{amssymb}
\usepackage{hyperref}
\usepackage{graphicx}
\usepackage{array}
\usepackage{siunitx}
\usepackage[usenames, dvipsnames]{color}
\usepackage{color}
\usepackage{epsfig}
\usepackage{longtable}
\usepackage[utf8]{luainputenc}

\usepackage{epsfig}
\usepackage{epstopdf}
\usepackage{xspace}
\usepackage{multicol}

\usepackage{makecell}
\usepackage{cite}

\usepackage{algorithm}
\usepackage{algorithmic}
\usepackage{float}
\usepackage{xcolor}
\usepackage{color}
\usepackage{float}

\usepackage{caption}
\usepackage{subcaption}	

\newcommand{\Smilei}{{\sc Smilei}\xspace}

\usepackage[margin=1.9cm]{geometry}
\usepackage{physics}

\usepackage{array}
\newcolumntype{L}{>{\centering\arraybackslash}m{3cm}}
\usepackage{multirow}

\hypersetup{%
   colorlinks=true,
   urlbordercolor={1 1 1},
   linkcolor={black},
   raiselinks=false,
citecolor=blue,
urlcolor=blue
}

\newcommand\blfootnote[1]{%
  \begingroup
  \renewcommand\thefootnote{}\footnote{#1}%
  \addtocounter{footnote}{-1}%
  \endgroup
}

\begin{document}

\title{\textbf{A Task Programming Implementation for the Particle in Cell Code \Smilei }}

\author{F. Massimo$^1$*, M. Lobet$^1$, J. Derouillat$^2$, A. Beck$^3$, \\G. Bouchard$^3$, M. Grech$^4$, F. Pérez$^4$, T. Vinci$^4$}
\date{\small{$^1$ Maison de la Simulation, CEA, France}\\
$^2$ Laboratoire des Sciences du Climat et de l'Environnement, CEA, France\\
$^3$ Laboratoire Leprince-Ringuet – École polytechnique, CNRS-IN2P3, France\\
$^4$ Laboratoire d’Utilisation des Lasers Intenses,  École polytechnique, CNRS, France}

\maketitle

\blfootnote{*Corresponding author. E-mail address: \href{mailto:francesco.massimo@cea.fr}{francesco.massimo@cea.fr}}

\begin{abstract}
An implementation of the electromagnetic Particle in Cell loop in the code \Smilei using task programming is presented.  Through OpenMP, the macro-particles operations are formulated in terms of tasks. This formulation allows asynchronous execution respecting the data dependencies of the macro-particle operations,  the most time-consuming part of the code in simulations of interest for plasma physics. Through some benchmarks it is shown that this formulation can help mitigating the load imbalance of these operations at the OpenMP thread level. The improvements in strong scaling for load-imbalanced physical cases are discussed.
\end{abstract}

\section{Introduction}
Electromagnetic Particle in Cell (PIC) \cite{BirdsallLangdon2004} codes represent state-of-the-art investigation tools for the plasma physics community and an important case study for optimization with novel high performance computing techniques. Indeed, the parallelization of PIC simulations becomes mandatory in many cases of physical interest,  which demand computing resources far beyond those obtainable with a single \textcolor{black}{CPU core}.

For a self-consistent treatment of the simulated phenomena,  at each iteration the PIC method advances the electromagnetic fields,  defined on a numerical grid, using macro-particles current densities as source terms in Maxwell's equations. Afterwards,  the macro-particles are advanced in the phase space using the electromagnetic fields as source terms of their \textcolor{black}{equations of motion}, project their current densities on the grid and the next PIC iteration can be executed. The mentioned operations on the macro-particles are the most time-consuming operations in a typical PIC simulation and are the first target of optimizations, as those presented in this work. 

\begin{figure}[!ht]
\centering
\includegraphics[scale=0.4]{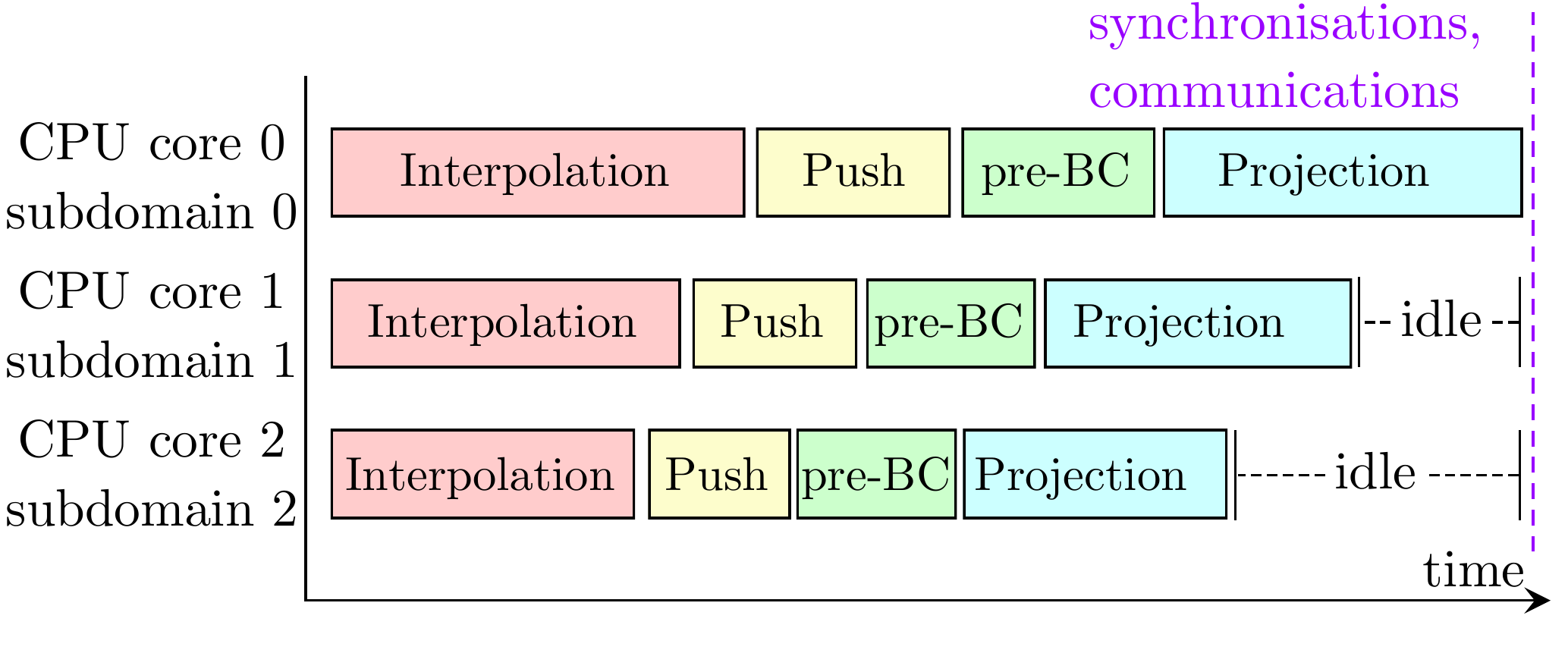}
\caption{Scheduling of macro-particle operations in a PIC loop iteration with a simplified case of 3 \textcolor{black}{CPU cores}. Each \textcolor{black}{CPU core} is assigned to a subdomain of the simulated physical space (macro-particles+fields). Each operator,  i.e. Interpolation, Push, macro-particle \textcolor{black}{preliminary operations for Boundary Conditions (pre-BC)} and current Projection, is applied to all macro-particles of the physical region assigned to the respective \textcolor{black}{CPU core}. The first \textcolor{black}{CPU core} is treating a physical zone with more particles than the other \textcolor{black}{CPU cores}, which remain idle while waiting for the synchronisation \textcolor{black}{and the following communications (e.g.  exchanges of macro-particles crossing domain borders, communication of fields at domain borders). }Note that the proportions between the PIC operators are not in scale: e.g. in a realistic example the time spent on \textcolor{black}{the pre-Boundary Conditions operations} would be much smaller than the time spent on Push.}
\label{fig:ToyPICcode}
\end{figure}

The standard parallelization of this algorithm consists in decomposing the physical domain in subdomains (including their fields and macro-particles) assigned to different computing units, \textcolor{black}{CPU cores} for the scope of this work.  Therefore, a load imbalance in the macro-particle operations occurs if they are unevenly distributed in space.  This load imbalance can be particularly detrimental in PIC codes, since at the end of each PIC loop iteration \textcolor{black}{at least one synchronization and fields/macro-particles communications at the subdomains borders are required.} Thus, the \textcolor{black}{CPU cores} which treated fewer macro-particles remain idle while the other ones treat physical regions with more particles, as depicted in Fig. \ref{fig:ToyPICcode} with a simplified model.

As long as the macro-particle distribution and thus the load is balanced, PIC codes have a good strong scaling and are often used to benchmark new machines/architectures.  However, a perfect load balance is not present in typical scientific applications, in particular when the strong scaling is challenged because in this case subdomains are taken as small as possible; as result, the simulation becomes sensitive to small scale imbalance, which has little impact on more coarsely parallelized cases. Indeed, physical set-ups starting with a load imbalance or with a load imbalance which develops during the simulation are easily encountered in PIC simulations of interest for the plasma physics community. Examples which often require large-scale simulations are laser wakefield acceleration \cite{Malka2002,Esarey2009,Malka2012}, interaction between a laser and a solid target \cite{Daido2012,Macchi2013,DiPiazza2012}, collisionless shocks in astrophysics \cite{Spitkovsky2008,Sironi2013,Plotnikov2018}.  Therefore,  for similar cases it is essential to improve the strong scaling, preventing the detrimental effects on performances of an uneven distribution of operations in a high number of \textcolor{black}{CPU cores}.

\begin{figure}[!ht]
\centering
\includegraphics[scale=0.4]{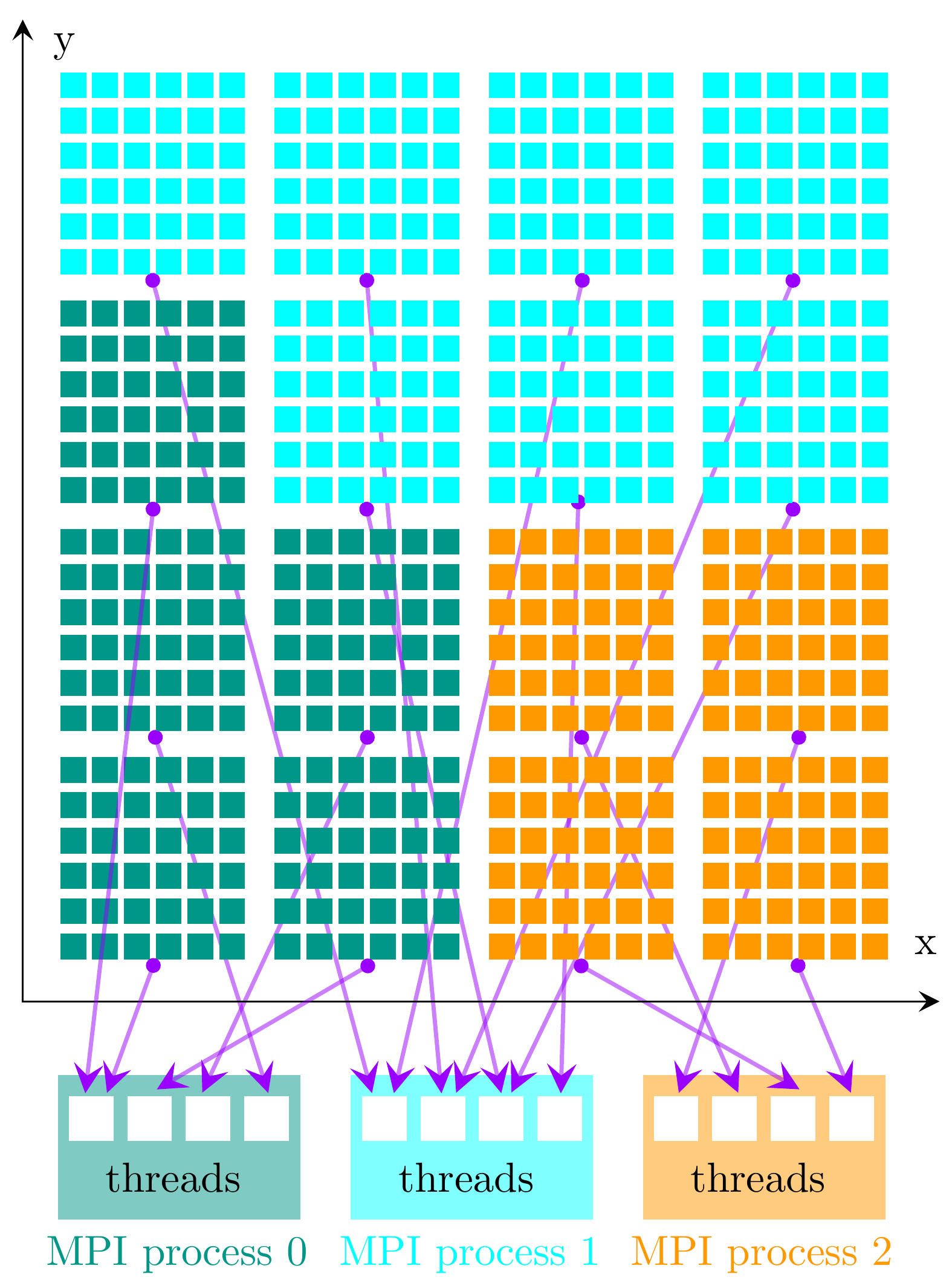}
\caption{Simple case of 2D domain decomposition with 4x4 patches in the $x$ and $y$ directions. Each patch has a size of 6x6 cells (small squares). Patches are colored according to the MPI process they are assigned to.  In the Figure 3 MPI processes are present, each with 4 OpenMP threads. In each \textcolor{black}{MPI patch collection} the patches are dynamically assigned to an OpenMP thread.}
\label{fig:PatchDecomposition}
\end{figure}

To reduce the effects of load imbalance, a hybrid MPI+OpenMP parallelization of the PIC method was implemented in some codes \cite{Germaschewski2016,Surmin2016,Vincenti2016,Smilei2018}.
It consists in decomposing the physical domain in small regions including multiple grid cells. These regions are called supercells / tiles / patches in the mentioned references.  In the following, only the term patches will be used.  Patches are then grouped in larger subdomains\textcolor{black}{, in this work called MPI patch collections, }each assigned to a MPI process.  This technique allows a dynamic load balancing algorithm exchanging patches between the \textcolor{black}{MPI patch collections},  whose form may adapt with irregular shapes to mitigate the load imbalance at the MPI level \cite{Germaschewski2016,Beck2016}.  Furthermore,  within \textcolor{black}{a MPI patch collection} the treatment of the patches is dynamically distributed between the OpenMP threads. This technique acts as a dynamic load balancing method at the finer level of the domain decomposition inside the \textcolor{black}{MPI patch collection} \cite{Smilei2018,Beck2019}.  An example of such decomposition is shown in Fig. \ref{fig:PatchDecomposition}.

\begin{figure}[!ht]
\centering
\includegraphics[scale=0.5]{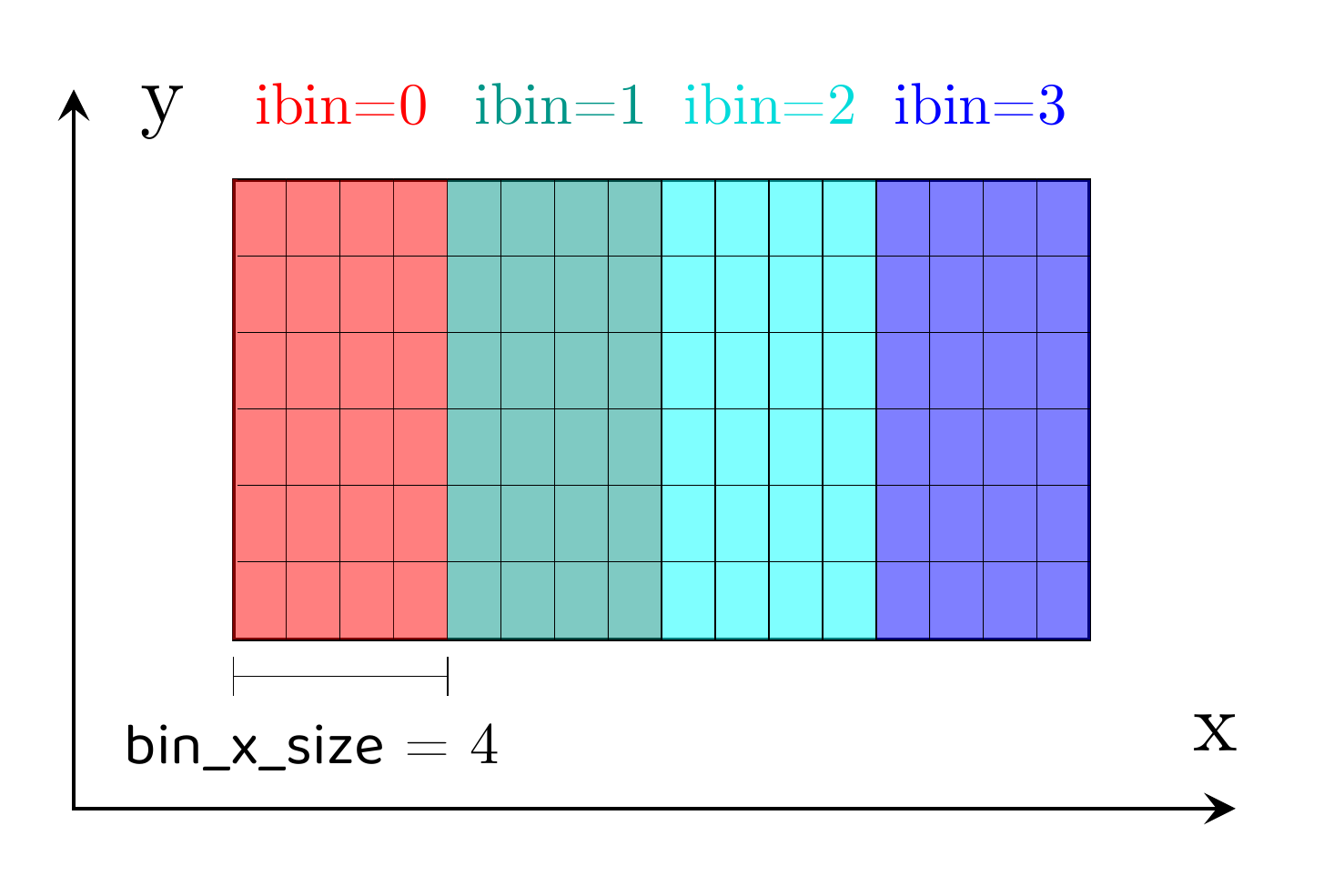}
\caption{Definition of bins in a patch. The depicted 2D patch's size is $16\times6$ cells in the $x$ and $y$ directions respectively.  In the Figure each bin has an $x$ extension equal to \texttt{\textcolor{black}{bin\_x\_size}} $=4$ cells in the $x$ direction. }
\label{fig:bin_definition}
\end{figure}

However,  in strongly imbalanced physical cases a load imbalance may remain at the OpenMP thread level,  i.e. at the patch level, where less-charged threads may have to wait for those charged with patches more populated of macro-particles (see Discussion in \cite{Derouillat2020}).

A technique of raising interest to mitigate the load imbalance at the OpenMP thread level in codes with hybrid parallelization is task programming \cite{Prat2018,ValeroLara2019,Klinkenberg2020}.  In this programming model, sequences of operations + their data scope are defined as tasks, which are generated and then asynchronously distributed at runtime among the OpenMP threads. The expression of data dependencies between different tasks is supported by most languages and API supporting task programming, e.g. OpenMP, used for this work \cite{openmp45}.

In a patch-based PIC code,  a task formulation of the algorithm allows more easily an even finer decomposition of the calculations within patches: these operations can be defined as sequences of tasks.  Namely, in this work a spatial finer domain decomposition within a patch called bin is defined.  \textcolor{black}{Thus, three spatial decompositions are used, with decreasing grain size: MPI patch collections, patches, bins (see Figs. \ref{fig:PatchDecomposition}, \ref{fig:bin_definition}). Each patch is composed by one or more bins, and \textcolor{black}{MPI patch collections include by definition} multiple patches. At each PIC loop iteration the macro-particles are sorted and assigned to macro-particle arrays corresponding to these levels of decomposition. }In this task formulation the operators acting on macro-particles of each bin (e.g. the current Projection operator) are defined as tasks.  With this paradigm, different operations on the bin macro-particles of the same patch are not bound to a particular OpenMP thread, but are executed by the first available thread.  This allows more flexibility in the scheduling of the operations and thus a more efficient distribution of the workload among the threads, provided that the overhead of tasks generation and scheduling is limited. Data dependencies between tasks ensure that the correct order of the PIC operations is respected. 

The implementation of this task-formulation in the open source PIC code \Smilei and its applications in some benchmarks are presented in this work,  \textcolor{black}{which is} organised as follows.

In the first section a quick review of the implementation of the PIC loop in \Smilei is presented in the context of its patch-based domain decomposition and hybrid MPI-OpenMP parallelization. In the second section the implementation of the task-based PIC loop in \Smilei \cite{Smilei2018} is presented. In the third section two small-scale simulations, a uniformly distributed plasma and a radiation pressure acceleration set-up, are used as case studies to show a comparison of the performances of the \Smilei versions without and with tasks.  The illustration of the scheduling of macro-particle operations in the second simulation is used to show the advantages in the scheduling of the version with tasks with load-imbalanced physical cases.  In the fourth section a more realistic simulation, nearer to those of practical interest, is used as benchmark, with small patches which allow a good load balancing at the MPI level and near to the minimum patch size. The performances and the strong scaling of the task-based version of the code are compared to those of the version without tasks.

The simulations of this work have been performed on the cluster Ruche of the Moulon Mesocentre \cite{Ruche}, on Intel Xeon Gold 6230 \textcolor{black}{(bi-socket Cascadelake node with 20 cores per socket, hyperthreading is not activated). }

\section{Parallelization without tasks}
\begin{figure}[ht]
  \centering
  \includegraphics[scale=0.35]{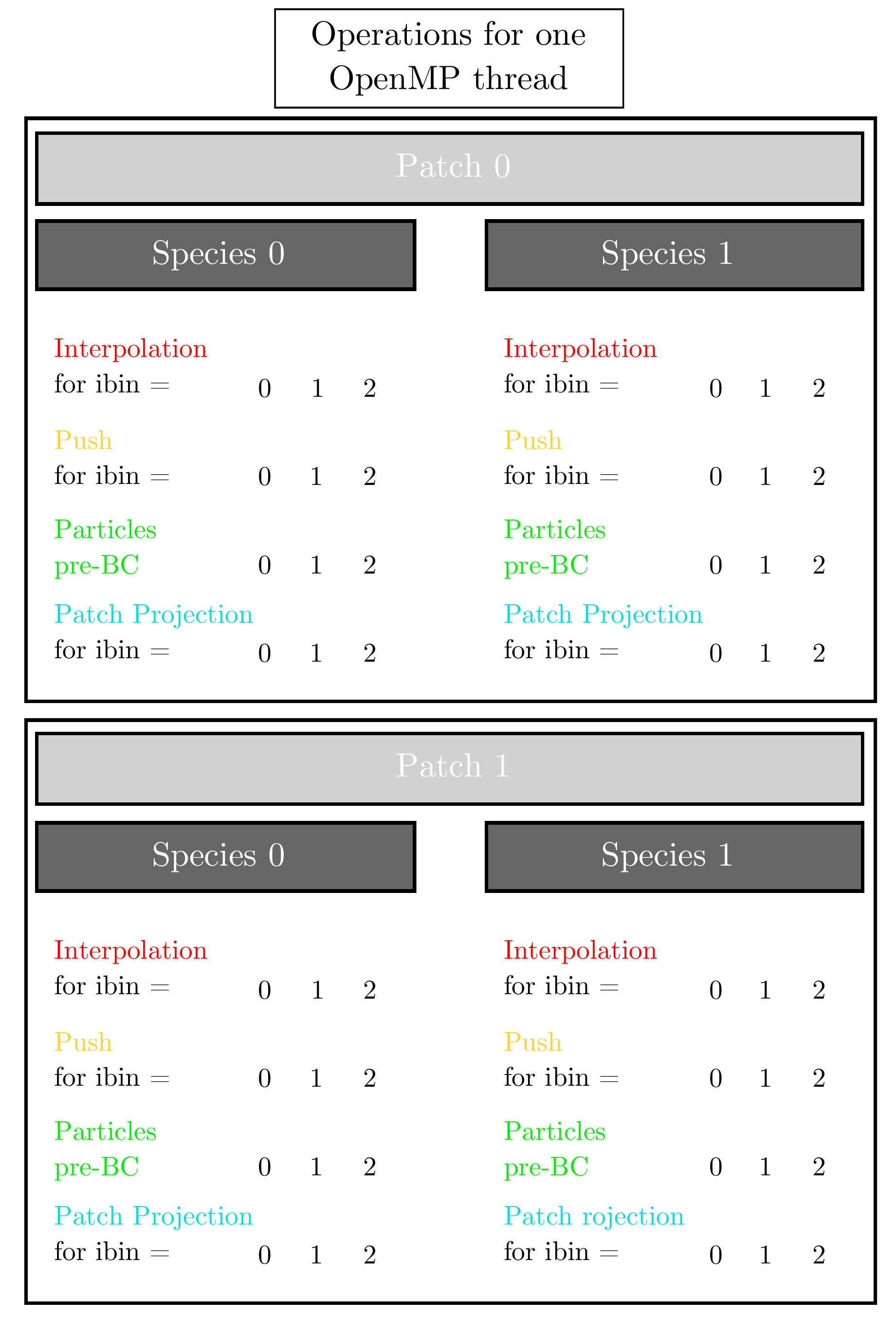}
  \caption{Macro-particle operations in \Smilei without tasks, depicting a case with two patches with two Species. \textcolor{black}{Note how all the operations corresponding to different species \texttt{ispec} belonging to the same patch \texttt{ipatch} are executed sequentially, by the same OpenMP thread.}}
\label{fig:NoTasksDiagram}
\end{figure}

In PIC codes,  the plasma distribution function in Vlasov equation is sampled with an ensemble of macro-particles with finite spatial extent.  The macro-particles move in the simulated domain representing the physical space, discretized with a numerical grid.  At each PIC algorithm iteration, for each macro-particle, the fields (and thus the Lorentz force) acting on the macro-particle are interpolated from the grid to the macro-particle position. Afterwards, Vlasov equation's characteristics, commonly called the \textcolor{black}{"equations of motion"} of the macro-particles, are solved to advance (or as more often said "push") the macro-particle position and momentum.  \textcolor{black}{Preliminary operations for Boundary Conditions are applied if the macro-particle reaches the borders of the physical domain (e.g. adding the macro-particle to the list of macro-particles to send to other patches or to delete). }Then, the current density of the particle is projected on the grid points associated \textcolor{black}{with its position}. The exact shape and extent of the density distribution of a given macro-particle depends on the chosen order of the spline functions used for interpolation and projection.  
After the execution of these operations on all the macro-particles, the total current density on the grid can be used to advance the electromagnetic fields solving Maxwell's equations.  Afterwards, a new PIC loop iteration can start.\\
\indent In the following, the mentioned operations involving macro-particles (Interpolation, Push, Projection, macro-particle \textcolor{black}{preliminary operations for Boundary Conditions or pre-BC}) are referred to as macro-particle operations.  These operations are always performed, independently of the parallelization and of diagnostic. The optimizations of the PIC algorithm in this work address these operations, the most time-consuming ones in most simulations of practical interest. \\
\indent In \Smilei, the parallelization of the described PIC method consists in decomposing the domain into \textcolor{black}{MPI patch collections, which are by definition} decomposed in small regions called patches, as shown in Fig. \ref{fig:PatchDecomposition} (see also \cite{Smilei2018}). Within each \textcolor{black}{MPI patch collection},  a loop on the patches (and on the macro-particle species within the patches) is performed at each iteration.  For all the macro-particles belonging to each patch \texttt{ipatch} and species \texttt{ispec}, the aforementioned macro-particles operations are executed.  As mentioned in the Introduction, to address load imbalance inside the MPI process, the patch loop can be parallelized with OpenMP with a dynamic scheduling, yielding the pseudo-code in Algorithm \ref{ompfor} (see also Figs. \ref{fig:PatchDecomposition},\ref{fig:NoTasksDiagram}). \textcolor{black}{Typically the best configuration for performances uses as many MPI processes as sockets and as many OpenMP threads per MPI process as CPU cores that are present in each socket. }For cache optimizations, the patch macro-particles are sorted in bins, i.e. group of particles corresponding to a decomposition of the patch physical space along the $x$ direction \textcolor{black}{(see Fig.\ref{fig:bin_definition})}. This decomposition will be particularly important in the task formulation of the algorithm, as detailed in the next section.\\
\indent Since a new PIC loop iteration cannot start before the current density projection of all patches, two synchronisations occur at the end of each iteration, corresponding to the \textcolor{black}{synchronization} between the OpenMP threads and the MPI processes.  \textcolor{black}{In this phase the exchanges of fields and macro-particles at the boundaries between patches and between MPI patch collections,  the patch exchange for load balancing, and the counting and sorting operations for SIMD vectorization are performed \cite{Beck2019}. The macro-particle Boundary Conditions (e.g. periodic, delete, send to other patch) are applied to the macro-particles flagged in the aforementioned preliminary operations for Boundary Conditions.} Load imbalances at these levels generate bottlenecks, as shown in Fig. \ref{fig:ToyPICcode}, that may be detrimental for strong scaling with a high number of \textcolor{black}{CPU cores}.\\
\indent The coarser load imbalance, between different MPI processes, is mitigated through the dynamic load balancing algorithm detailed in \cite{Beck2016}. With a frequency chosen by the user, the load imbalance is evaluated and patches are exchanged between MPI processes to have a MPI-wise macro-particle distribution as uniform as possible.  This strategy changes the shape of the \textcolor{black}{MPI patch collections}, which generally deviate from a ``rectangular" shape.  

\begin{algorithm}[H]
\caption{Macro-particle operations in a PIC loop iteration,  version \textcolor{red}{``Tasks OFF"}.  Each operation call, e.g. Interpolation, is executed on all the macro-particles of the combination [\texttt{ipatch},\texttt{ispec},\texttt{ibin}]. Note that the \textcolor{magenta}{\#pragma omp for schedule(dynamic)} dynamically distributes the \texttt{ipatch} iterations to the OpenMP threads at runtime.}\label{ompfor}
\begin{algorithmic}
\STATE \textcolor{magenta}{\#pragma omp for schedule(dynamic)}
\FOR{\texttt{ipatch=0,$\rm{N_{patches}-1}$}}
\FOR{\texttt{ispec=0,$\rm{N_{species}-1}$}}
\FOR{\texttt{ibin=0,$\rm{N_{bins}-1}$}}
\STATE \textcolor{violet}{\texttt{Interpolation}}
\ENDFOR
\FOR{\texttt{ibin=0,$\rm{N_{bins}-1}$}}
\STATE \textcolor{violet}{\texttt{Push}}
\ENDFOR
\FOR{\texttt{ibin=0,$\rm{N_{bins}-1}$}}
\STATE \textcolor{violet}{\texttt{Preliminary operations for Boundary Conditions}}
\ENDFOR
\FOR{\texttt{ibin=0,$\rm{N_{bins}-1}$}}
\STATE\textcolor{violet}{ \texttt{Projection of current density into ipatch grid}}
\ENDFOR
\ENDFOR
\ENDFOR
\end{algorithmic}
\end{algorithm}

The patch-level load imbalance is normally mitigated by the dynamic distribution of patches among the threads of a MPI process.  However, two constraints of this approach may cause bottlenecks when the workload on different patches is unevenly distributed. As described by Algorithm \ref{ompfor} and depicted in Fig.\ref{fig:NoTasksDiagram}, at each patch of an \textcolor{black}{MPI patch collection} corresponds a thread that is dynamically assigned at runtime.  Namely, this implies the following constraints in the scheduling of macro-particle operations: 
\begin{itemize}
\item all the species within a patch are treated by the same OpenMP thread;
\item this thread must also execute all the involved operations for that patch and its species. 
\end{itemize}
This version of the algorithm parallelization is referred to in the following as \textcolor{red}{Tasks OFF}. 

These constraints may generate bottlenecks when the number of OpenMP threads is near to the number of patches, a situation that can occur when many \textcolor{black}{CPU cores} are used and/or the dynamic load balancing algorithm assigns few patches to a certain MPI process (see e.g. the Discussion section in \cite{Derouillat2020}).  In this situation the strong scaling worsens, since a balanced distribution of the workload in patches among the threads is not possible.  Reducing the size of the patches may help,  but a patch cannot be smaller than a certain size, determined by the number of grid points involved by the interpolation/projection spline function, i.e. the extent of the macro-particles. 

Therefore, it is useful to introduce additional flexibility in the distribution of macro-particles operations among the OpenMP threads, to remove the mentioned constraints of version \textcolor{red}{Tasks OFF}. In the next section a task formulation of these operations in \Smilei is presented. It is shown in the following sections that the increased flexibility in scheduling improves the strong scaling and reduces the simulation time in load-imbalanced cases.

\section{Parallelization with OpenMP task}\label{taskversion}
The implementation of \Smilei with task programming presented in this section, referred to as version \textcolor{blue}{Tasks ON}, is not the only possible one in an electromagnetic PIC code (see e.g.  \cite{Saez2015,Saez2016,Guidotti2021}). One of the advantages of this formulation is its compatibility with the patch-based domain decomposition and load balancing strategies at MPI level described in the previous section.  For a small review of the OpenMP task concepts of interest for this work,  please refer to the OpenMP specification version 4.5 \cite{openmp45}, in particular the \texttt{task}, \texttt{taskwait}, \texttt{taskgroup} constructs and the \texttt{depend} clause.

As first level of ``taskification", each iteration of the outer double \texttt{for} loop in Algorithm \ref{ompfor} can be formulated as a task.  Conceptually the Interpolation, Push and \textcolor{black}{pre-BC} operations on the macro-particles in different combinations [\texttt{ipatch},\texttt{ispec}] can be independent and do not have a risk of race conditions if a separate grid for a given [\texttt{ipatch},\texttt{ispec}] combination is used and auxiliary variables are defined as local variables. Indeed, interpolation is a read operation from the grid and a write operation on each of the macro-particle buffers (macro-particle position index, interpolated fields, ...). Keeping these buffers as local variables ensures that no race condition is encountered in asynchronous execution of the Interpolation, although special care is needed to avoid memory leaks. The Push operation is a read operation on the independent macro-particle buffers and a write operation on the independent macro-particles positions and momenta. The \textcolor{black}{preliminary operations for Boundary Conditions} are write operations on the independent macro-particles positions and momenta. However, in Algorithm \ref{ompfor} the Projection operation yields a crucial risk of data race if particles from the same \texttt{ipatch} but different \texttt{ispec} write their corresponding current/charge density on the same grid point.  Using \texttt{atomic} operations in the Projection to avoid race conditions would have detrimental effects on the performances. Thus, a special modification of the projection operation has been implemented,  as detailed in the following.  It consists in a preliminary projection operation which is local for a given combination [\texttt{ipatch},\texttt{ispec}]. For this purpose, for each [\texttt{ipatch}, \texttt{ispec}, \texttt{ibin}], dedicated local grids are defined.  Each of these local subgrids has the same number of cells as the physical space corresponding to a bin, and the necessary ghost cells at their borders.

As second step of the \Smilei task formulation,  tasks are defined at the level of decomposition of bins (see Fig. \ref{fig:bin_definition}). Each bin has an extension of \texttt{\textcolor{black}{bin\_x\_size}} cells in the $x$ direction.  With this finer grain of parallelism, the parameter \texttt{\textcolor{black}{bin\_x\_size}} can be used to choose the task granularity, whose optimal value may vary depending on the physical set-up.

Inside the aforementioned [\texttt{ipatch},\texttt{ispec}] tasks, the macro-particle operations on macro-particles belonging to different bins can be formulated as subtasks, with the same data race risks described above.  Thus,  differently from the Algorithm \ref{ompfor},  in principle the macro-particle operations of each combination [\texttt{ipatch},\texttt{ispec},\texttt{ibin}] can be performed asynchronously on different threads as tasks, provided that the sequence Interpolation-Push-\textcolor{black}{(pre-BC)}-Projection of the selected [\texttt{ipatch},\texttt{ispec},\texttt{ibin}] combination is respected and race conditions are avoided.  The sequence order can be ensured through the \texttt{depend} clause of the \texttt{omp task} directive. To avoid race conditions between macro-particles belonging to different \texttt{ispec} on the same \texttt{ipatch} projecting on the same grid points, each [\texttt{ipatch},\texttt{ispec},\texttt{ibin}] combination projects on its local sub-grid corresponding to the physical space of [\texttt{ipatch},\texttt{ibin}] (see Fig. \ref{fig:bin_definition}) and the necessary ghost cells.   

After the macro-particle operations on the patch \texttt{ipatch} are executed,  a reduction of the subgrids current densities can be executed on the main grid, whose fields are used as source terms to solve Maxwell's equations.  The current densities in the mentioned local subgrids corresponding to the bins are added to the patch grid densities with the appropriate shift (an integer multiple of \texttt{\textcolor{black}{bin\_x\_size}} cells). This operation is executed for each patch, species and bin.

It is important to note that, provided that the mentioned task dependencies are respected, different macro-particle operations on different patches or species or bins may be distributed on other OpenMP threads by the task scheduler, without \textcolor{black}{binding} the operation of the particles in a given [\texttt{ipatch},\texttt{ispec},\texttt{(ibin=0, $\rm{N_{bins}}$-1)}] to a particular (dynamically assigned) OpenMP thread as in Algorithm \ref{ompfor}.  
For the reduction operation of a given [\texttt{ipatch},\texttt{ispec}] combination,  it was chosen to set it as dependent on the projection of all the particles belonging to the combination [\texttt{ipatch},\texttt{ispec}] (all the bins) and on the same reduction executed on the combination [\texttt{ipatch},\texttt{ispec-1}] (all the bins). The loop on the bins of a given [\texttt{ipatch},\texttt{ispec}] combination is performed serially. Since the macroparticles are typically much more numerous than the grid points and given the number of operations involved in the macro-particles operations, the time spent \textcolor{black}{on} this additional reduction step is normally negligible compared to the time spent on macro-particle operations (see also Fig. \ref{fig:task_tracing2D} in the next section).
 
\begin{figure}[!ht]
  \centering
  \includegraphics[scale=0.4]{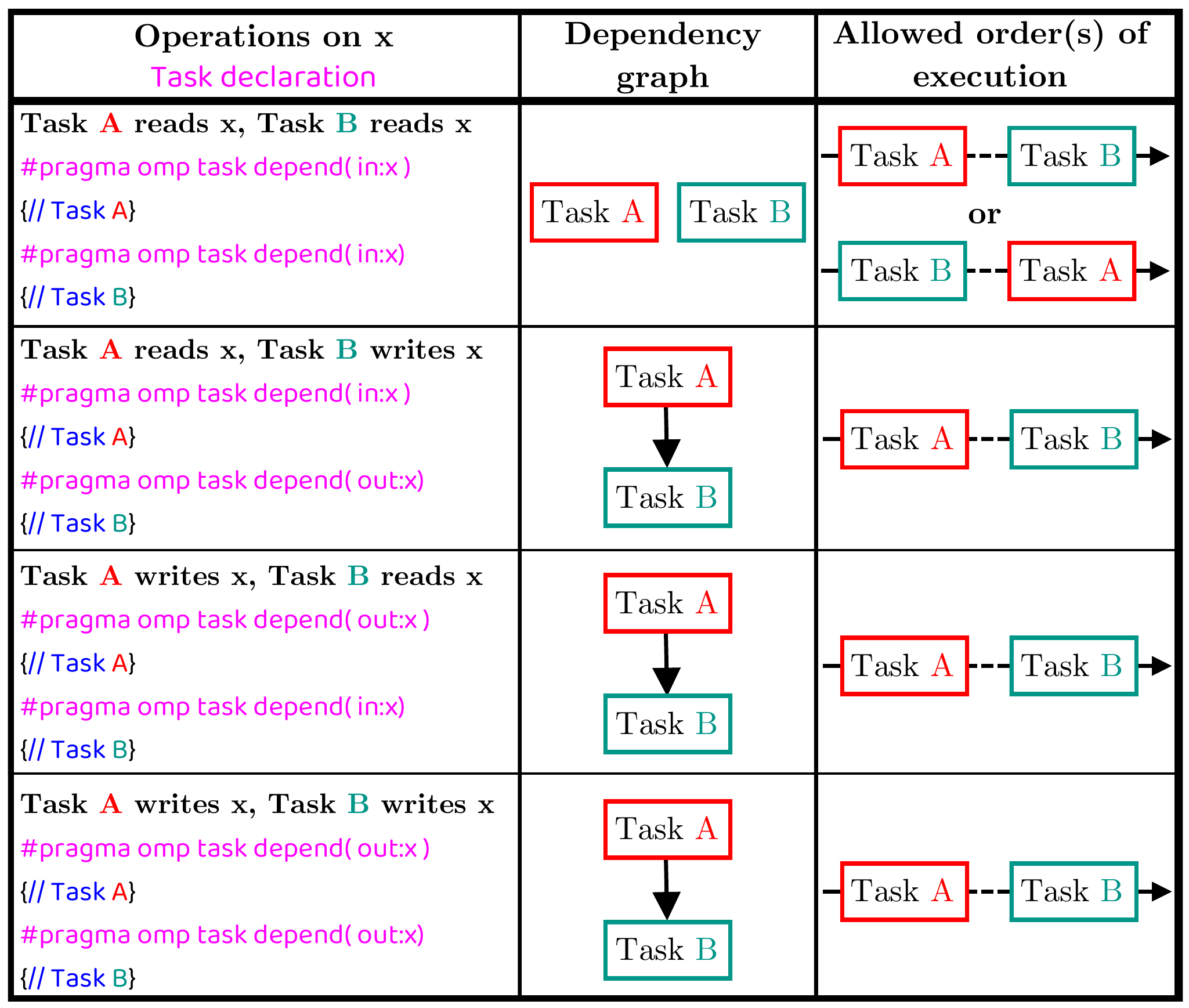}
  \caption{Dependency graph and allowed order(s) of executions for tasks A and B according to the dependency type specified on their \texttt{\textcolor{magenta}{depend}} clause. The variable \texttt{x} (which may be an array element) in the dependency list may be a variable treated inside the tasks or simply a variable defined to be used as tag for dependencies. The OpenMP runtime scheduler will assign the tasks to the available threads respecting the allowed execution order(s).
In the dependency graph column the arrow symbolizes a dependency.  In the right column, the arrow represents the time axis. The dashed line is used to mean that in that time interval other tasks may be executed.}
\label{task_depend_clause}
\end{figure}

\begin{figure}[!ht]
  \centering
  \includegraphics[scale=0.35]{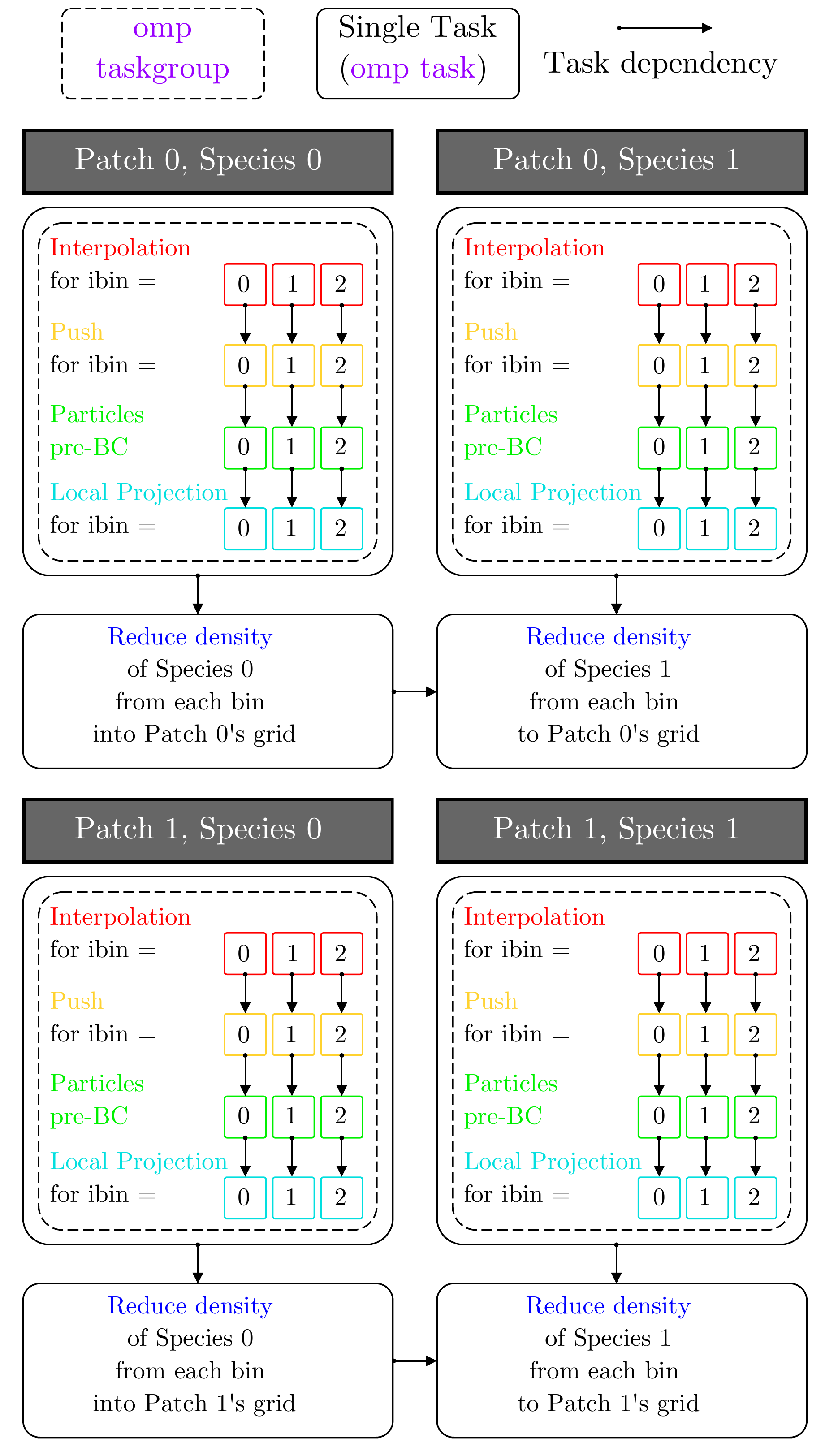}
  \caption{Task dependency graph of macro-particle operations in \Smilei, depicting a case with two patches with two Species. Note how the tasks corresponding to different [\texttt{ipatch,ispec,ibin}] combinations (and different operators) are independent and thus can be assigned to any of the available threads respecting the dependencies.  The \texttt{taskgroup} construct ensures that the enclosed tasks and their child tasks are completed. \textcolor{black}{Please note that the grey boxes refer to the operations for the particles in the combination [\texttt{ipatch,ispec}]. The rectangles/squares with solid border and smooth vertices representing a task (see legend in the top part of the figure) are colored according to the color-coded operators in the Figure. } }
\label{fig:task_dependency_graph}
\end{figure}

\textcolor{black}{The described task formulation is depicted in Algorithm \ref{omptask}, where the \textcolor{magenta}{\texttt{task}} clauses refer to the generation of a task. The OpenMP thread which will execute that task and the moment when this will occur are dynamically determined by the OpenMP runtime scheduler.  The tasks appearing in Algorithm \ref{omptask} have a dependence type \textcolor{magenta}{\texttt{in}} and/or \textcolor{magenta}{\texttt{out}}, followed by a list of variables (which may be array elements). This clause specifies the dependencies between tasks. To understand these clauses, intuitively a \textcolor{magenta}{\texttt{depend(out:x)}} clause in a \textcolor{magenta}{\texttt{task}} construct treats the task as if the variable \textcolor{magenta}{\texttt{x}} is written by the task, while a \textcolor{magenta}{\texttt{depend(in:x)}} clause treats the task as if the variable \textcolor{magenta}{\texttt{x}} is read by the task. Thus, the OpenMP scheduler constructs the dependence relations of the new task based on the previously generated tasks and on the requirement that the variable \textcolor{magenta}{\texttt{x}} should not be subject to data races.  Therefore, following the OpenMP 4.5 specification \cite{openmp45} for the \textcolor{magenta}{\texttt{in}} dependence-type of a task B, if at least one of the listed variables is in common with those of a previously generated task A with an \textcolor{magenta}{\texttt{out}} dependence-type, then task B will depend on (and therefore will be executed after) task A.  For the \textcolor{magenta}{\texttt{out}} dependence-type of a task B, if at least one of the listed variables is in common with a previously generated task A with dependence type \textcolor{magenta}{\texttt{in}} or \textcolor{magenta}{\texttt{out}}, then then task B will depend on (and therefore will be executed after) task A. These rules for the task execution order are depicted in Fig.  \ref{task_depend_clause}.} The dependencies expressed in Algorithm \ref{omptask} are also depicted in Fig. \ref{fig:task_dependency_graph}. \textcolor{black}{Remembering that each task depicted in that Figure can be executed virtually by any available OpenMP thread,  a comparison with Fig. \ref{fig:NoTasksDiagram} highlights the already mentioned constraints that are present in the version without tasks when the operations must be assigned to the threads. }

It is worth noting that even if in a version without tasks a triple loop on \texttt{ipatch}, \texttt{ispec}, \texttt{ibin} was used to have a finer grain decomposition for the scheduling of particle operations,  it would not be straightforward to respect the dependencies with an OpenMP directive not using tasks.  Besides, without a local buffer projection and subsequent reduction, a risk of race condition would occur with the projection operation. Future work will also show the performances of a version with a double for loop with the \texttt{collapse(2)} clause to assign to different threads the operations on different \texttt{ipatch}, \texttt{ispec}.

A final consideration on this version with tasks is the overhead in memory added, \textcolor{black}{compared to the version without tasks}.  The version in Algorithm \ref{ompfor} uses \textcolor{black}{buffer arrays} for quantities like spatial indices of the macro-particle on the grid (3 integers in 3D), \textcolor{black}{displacement of the macro-particle position referred to the borders of the cells associated to those indices} (3 real numbers in 3D),  \textcolor{black}{the electromagnetic fields interpolated at each macro-particle's position (6 real numbers, one for each component of the electromagnetic field)}. These quantities are computed by the Interpolation operation and used by the Pusher and Projection operation \cite{BirdsallLangdon2004}, \textcolor{black}{reason why the mentioned arrays are used to store the data to use again later}.  \textcolor{black}{There can be a maximum of \texttt{Nthreads} OpenMP threads executing the operations on \texttt{Nthreads} patches at a given time. Thus, to avoid risks of race conditions, }the memory needed for the mentioned buffer arrays corresponds to that of \texttt{Nthreads} arrays with \textcolor{black}{a number of elements corresponding to the number of macro-particles treated by that thread in the involved operations, \texttt{Nmacro-particles}.}  Each time a thread starts to treat a species in a patch assigned to it,  \textcolor{black}{the number of macro-particles} \texttt{Nmacro-particles} in the considered [\texttt{ipatch},\texttt{ispec}] combination is computed and \textcolor{black}{the thread's buffer array is dynamically resized.} This approach does not work in the version with tasks in Algorithm \ref{omptask}, since each of the \texttt{Nthreads} may execute the operations of an operator (e.g. Interpolator) on a certain [\texttt{ipatch},\texttt{ispec},\texttt{ibin}] combination and immediately afterwards \textcolor{black}{execute the same operations} on another combination \textcolor{black}{[\texttt{ipatch},\texttt{ispec},\texttt{ibin}]}, or \textcolor{black}{may execute} the operations of another operator on any combination \textcolor{black}{[\texttt{ipatch},\texttt{ispec},\texttt{ibin}]. Therefore the same [\texttt{ipatch},\texttt{ispec},\texttt{ibin}] combination may be treated by different threads for different operators}. Thus, instead of \texttt{Nthreads} \textcolor{black}{buffer} arrays, the number of buffers is set to \texttt{Npatch}$\times$\texttt{Nspecies}. Each \textcolor{black}{buffer array} has an index that is given by \texttt{ipatch}$\times$\texttt{Nspecies}$+$\texttt{ispec}, where [\texttt{ipatch},\texttt{ispec}] are the indices of the patch and species associated to it. Resizing each of these arrays to the respective sizes \texttt{Nmacro-particles} and leaving them with that size would easily generate a memory leak. Thus, to reduce the memory overhead, the size of these arrays is set to 1 when they are no more necessary, i.e. when the projection operation for the combination [\texttt{ipatch},\texttt{ispec},\texttt{ibin}] has been completed \textcolor{black}{and no further operations are needed on the macro-particles in that region of space until the next PIC loop iteration}.\\
\indent In addition,  as mentioned above,  \textcolor{black}{the task version needs to define for each [\texttt{ipatch},\texttt{ispec}, \texttt{ibin}] combination a local sub-grid corresponding to the physical space of a bin and the necessary ghost cells at its border.  As explained above in this section, these additional sub-grids are necessary to avoid race conditions in the current/charge density projections.}

\begin{algorithm}[H]
\caption{Macro-particle operations in a PIC loop iteration,  version \textcolor{blue}{``Tasks ON"}.  Note that the second dependency of the second task ensures that the current density reduction on a given patch \texttt{ipatch} is performed in sequence for all the species (see also Fig.\ref{fig:task_dependency_graph}).  Each operation call, e.g. Interpolation, is executed on all the macro-particles of the combination [\texttt{ipatch},\texttt{ispec},\texttt{ibin}].  \textcolor{black}{Please refer to section \ref{taskversion} and to the OpenMP 4.5 specification \cite{openmp45} for the definition of the \textcolor{magenta}{\texttt{depend(in:...)}} and \textcolor{magenta}{\texttt{depend(out:...)}} clauses. }}\label{omptask}
\begin{algorithmic}
\STATE \textcolor{magenta}{\#pragma omp single}
\STATE  \{
\FOR{\texttt{ipatch=0,$\rm{N_{patches}-1}$}}
\FOR{\texttt{ispec=0,$\rm{N_{species}-1}$}}
\STATE \textcolor{magenta}{\#pragma omp task firstprivate(ipatch,ispec)}
\STATE \textcolor{magenta}{depend(out:has\_done\_dynamics[ipatch][ispec]}
\STATE  \textcolor{violet}{\texttt{dynamics(ipatch,ispec)}}
\STATE \textcolor{magenta}{\#pragma omp task firstprivate(ipatch,ispec)}
\STATE \textcolor{magenta}{depend(in:has\_done\_dynamics[ipatch][ispec])}
\STATE \textcolor{magenta}{depend(out:has\_reduced\_densities[ipatch])}
    \STATE \{
        \hskip2.0em\FOR{\texttt{ibin=0,$\rm{N_{bins}-1}$}}
            \STATE  \textcolor{violet}{\texttt{Copy current density of local sub-grid}}
             \STATE  \textcolor{violet}{\texttt{into ipatch grid}}
        \ENDFOR
    \STATE \} \textcolor{blue}{$\slash\slash$ end ipatch-ispec density reduction task}
\ENDFOR
\ENDFOR
\STATE  \} \textcolor{blue}{$\slash\slash$ end omp single}
\STATE
\STATE def \textcolor{violet}{\texttt{dynamics(ipatch,ispec)}}:
\STATE  \{
\STATE \textcolor{magenta}{\#pragma omp taskgroup}
\STATE  \{
\FOR{\texttt{ibin=0,$\rm{N_{bins}-1}$}}
\STATE \textcolor{magenta}{\#pragma omp task firstprivate(ibin)}
\STATE \textcolor{magenta}{depend(out:has\_interpolated[ibin])}
\STATE \textcolor{violet}{\texttt{Interpolation}}
\ENDFOR
\STATE 
\FOR{\texttt{ibin=0,$\rm{N_{bins}-1}$}}
\STATE \textcolor{magenta}{\#pragma omp task firstprivate(ibin)}
\STATE \textcolor{magenta}{depend(in:has\_interpolated[ibin]) }
\STATE \textcolor{magenta}{depend(out:has\_pushed[ibin])}
\STATE \textcolor{violet}{\texttt{Push}}
\ENDFOR
\STATE 
\FOR{\texttt{ibin=0,$\rm{N_{bins}-1}$}}
\STATE \textcolor{magenta}{\#pragma omp task firstprivate(ibin)}
\STATE \textcolor{magenta}{depend(in:has\_pushed[ibin])}
\STATE \textcolor{magenta}{depend(out:has\_done\_bc[ibin])}
\STATE \textcolor{violet}{\texttt{Preliminary operations for Boundary Conditions}}
\ENDFOR
\STATE 
\FOR{\texttt{ibin=0,$\rm{N_{bins}-1}$}}
\STATE \textcolor{magenta}{\#pragma omp task firstprivate(ibin)}
\STATE \textcolor{magenta}{depend(in:has\_done\_bc[ibin])}
\STATE \textcolor{violet}{\texttt{Projection of current density}}
\STATE \textcolor{violet}{\texttt{into local ipatch,ispec,ibin subgrid}}
\ENDFOR
\STATE  \} \textcolor{blue}{$\slash\slash$ end omp taskgroup}
\STATE  \} \textcolor{blue}{$\slash\slash$ end dynamics}
\end{algorithmic}
\end{algorithm}

\begin{flushleft}
\section{Basic performance benchmarks}\label{sec:four_code_versions}
\end{flushleft}

To illustrate the advantages of the task formulation in the scheduling of macro-particle operations with a toy example,  the performances of tasks with two simple physical cases are reported in this section. Afterwards, the scheduling of macro-particle operations will be displayed for two representative cases drawn from the considered performance scans. \textcolor{black}{For the sake of simplicity, these simplified simulations have few large patches compared to the number of CPU cores. With many small patches, }both the MPI load balancing and the OpenMP dynamic scheduling would work more efficiently in reducing the idle-times of \textcolor{black}{CPU cores} caused by load imbalances in the macro-particle distribution. 

In the following, two cases studies are presented, where the two mentioned versions of the algorithm are used (without and with tasks, the already defined \textcolor{red}{``Tasks OFF"} and \textcolor{blue}{``Tasks ON"}), varying the number of total \textcolor{black}{CPU cores} and the bin size along the  $x$ direction \texttt{\textcolor{black}{bin\_x\_size}} $=[1,2,4,8,16,32,64]$. The times reported in the plots represent the time spent on the macro-particle operations: Interpolation, Push, Particles BC, Projection, and Density Reduction if present.

The first case study is a small simulation of a uniform plasma, with perfect load balancing during all the simulation. In this 2D simulation, a uniform preionized plasma of density $1$ $n_c$ is initialized with electron temperature $T_e=100\thinspace m_ec^2$ and ion temperature $T_i=10\thinspace m_ec^2$, extending up to the $x$ and $y$ borders of the simulation window.  Both electron and ion macro-particles are initialized with \textcolor{black}{a} regular distribution along the cell and 36 macro-particles per cell. The grid has size $L_x=512\Delta x$ and $L_y=1024\Delta y$, sampled with resolution $\Delta x= \Delta y=0.22$ $c/\omega_r$ (half of the Debye length), where $\omega_r$ can be an arbitrary reference angular frequency.  The simulation domain is decomposed in $8\times8$ patches of size $64\times128$ cells. A spline of order 2 is used for Interpolation and Projection. Periodic boundary conditions are used at the borders of both directions, for both particles and electromagnetic fields.

\begin{figure}[!ht]
\centering
\includegraphics[scale=0.6]{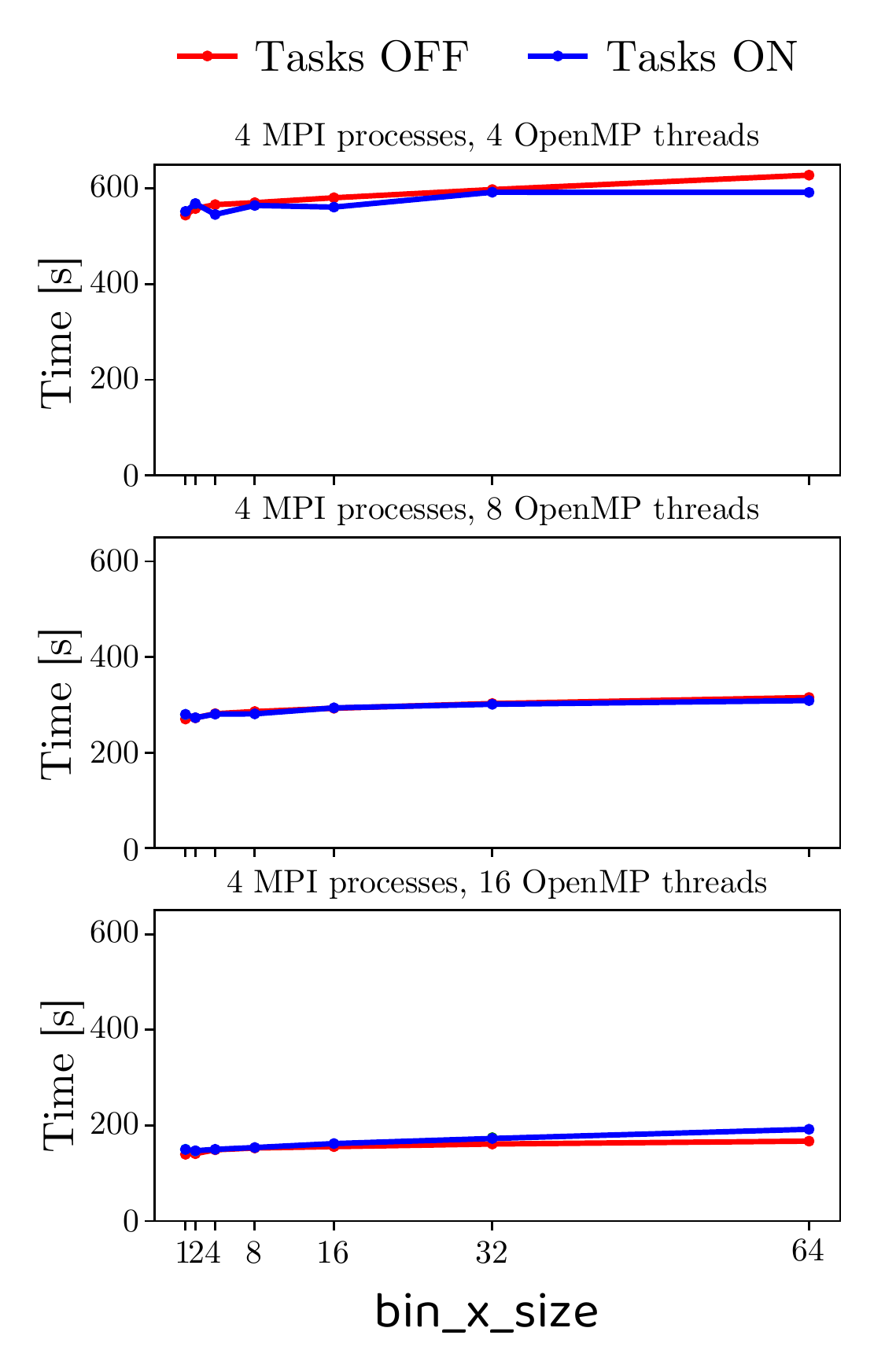}
\caption{Scan varying \texttt{\textcolor{black}{bin\_x\_size}} of the 2D uniform plasma benchmark, \textcolor{black}{with and without tasks}. The reported times for macro-particle operations are measured for 1500 iterations and averaged over the MPI processes. }
\label{fig:uniform_plasma2D}
\end{figure}

From Fig. \ref{fig:uniform_plasma2D} it is inferred that in this set-up and with the considered number of \textcolor{black}{CPU cores}, both versions have nearly the same strong scaling in the time spent in macro-particle operations.  \textcolor{black}{As expected for a balanced case, the version with tasks and the version without tasks spend the same time on macro-particle operations. The amount of operations to execute is approximately the same for each patch, thus the idle times (like those in Fig. \ref{fig:ToyPICcode}) are minimized. The additional degrees of freedom of the task formulation in the assignment of the operations to the OpenMP threads do not yield an advantage compared to a dynamically scheduled \texttt{omp for} loop. It is worth noting also that the overhead given by the density reductions in the task formulation does not seem to have sensitive effects on the performances in this selected benchmark. }

\begin{figure}[!ht]
\centering
\includegraphics[scale=0.6]{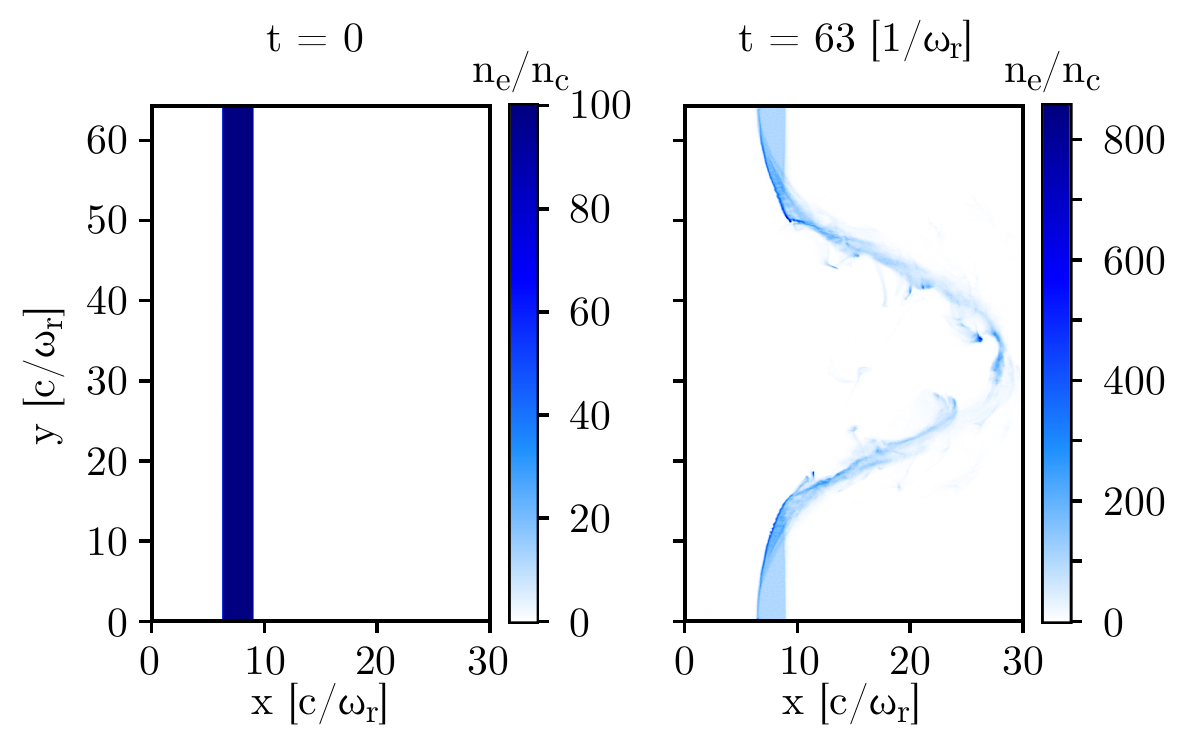}
\caption{Electron density $n_e$ divided by the critical density $n_c$ in the 2D radiation pressure benchmark at 0 (left) and 1500 iterations (right).  }
\label{fig:radiation_pressure2D_rho}
\end{figure}

The second case study is a small-scale simulation of a short slab of pre-ionized plasma irradiated by a laser. \textcolor{black}{In this case the motion of the electron species and the motion of the heavier ion species have some differences, maintaining a load imbalance even between the patches of a MPI process. }At $t=0$ the macro-particle load is evenly distributed along the MPI processes and dynamic load balancing is activated.  To illustrate the load imbalance of the electron distribution in this physical case, Figure \ref{fig:radiation_pressure2D_rho} displays the electron density at 0 and 1500 iterations. The ion density (not shown) has a similar unbalanced distribution.

In this 2D simulation, a slab of preionized plasma of uniform density $100$ $n_c$, thickness $0.44$ $c/\omega_r$ is initialized at $x=\lambda_r$, extending up to the $y$ borders of the simulation window.  The plasma is sampled with electron and ion macro-particles, both initialized with \textcolor{black}{a} regular distribution and 64 macro-particles per cell. Initially the ions are cold and the electrons have an initial temperature of $0.001$ $m_ec$. The grid has size $L_x=512\Delta x$ and $L_y=1024\Delta y$, sampled with resolution $\Delta x= \Delta y=\lambda_r/100$.  The simulation domain is decomposed in $8\times8$ patches of size $64\times128$ cells. The integration timestep is set to $\Delta t=(\lambda_r/c)/150$. A transversely Gaussian laser with peak normalized $a_0=150$,  arbitrary carrier frequency $\lambda_r$, waist $w_0=2\lambda_r$, constant temporal profile starts to be injected from the left $x$ border of the window and focused at $[x,y]=[10\lambda_r,L_y/2]$, triggering radiation pressure acceleration of the particles in the plasma slab.  Dynamic load balancing at the MPI level is activated at the start of the simulation and every 20 iterations.  A spline of order 2 is used for Interpolation and Projection. Periodic boundary conditions for particles and fields are used at the borders in the $y$ direction, while reflective and Silver-Muller conditions are used at the borders in the $x$ direction for the particles and the electromagnetic fields respectively. 

\begin{figure}[!ht]
\centering
\includegraphics[scale=0.6]{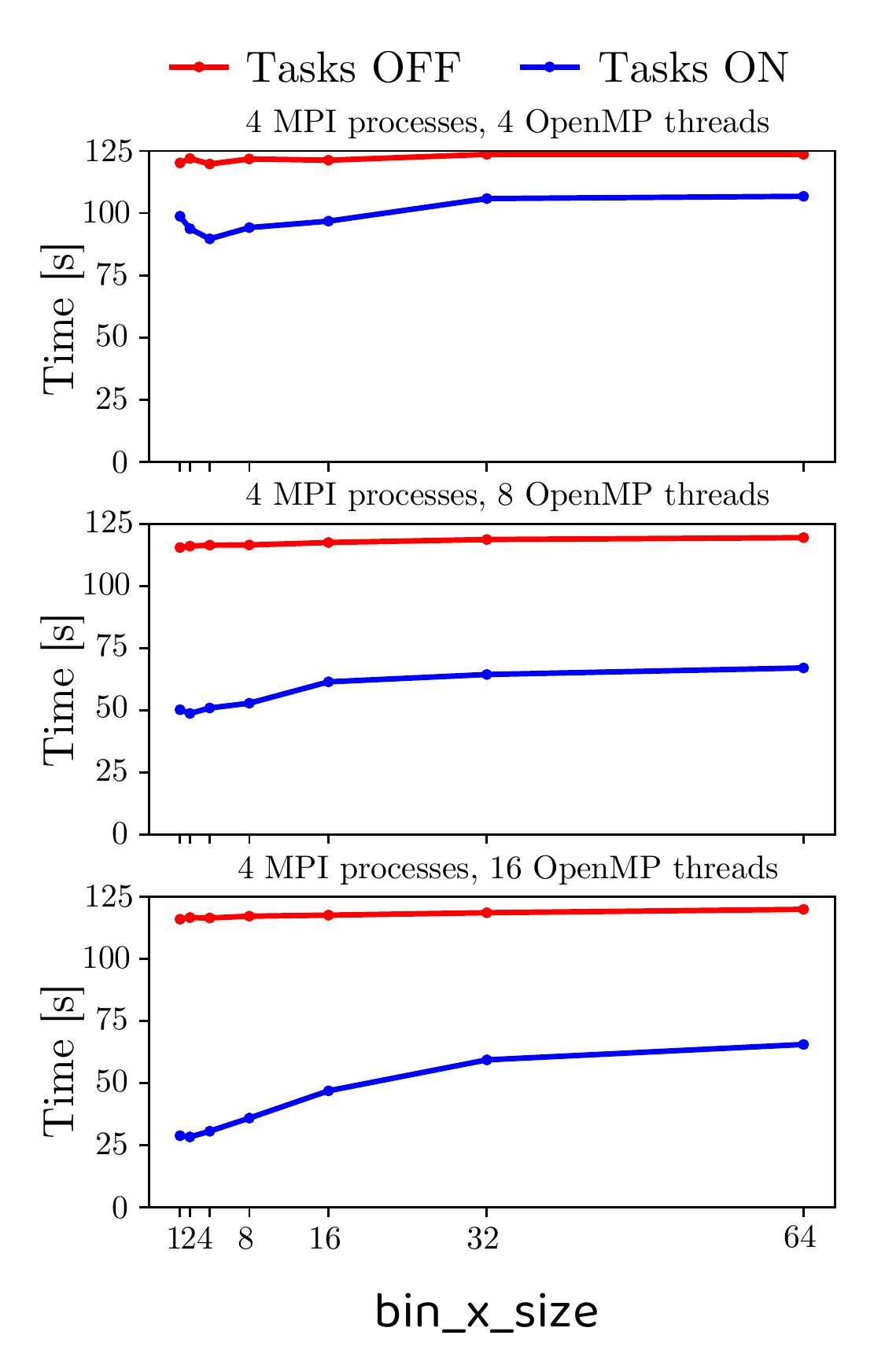}
\caption{Scan varying \texttt{\textcolor{black}{bin\_x\_size}} of the 2D radiation pressure benchmark,  \textcolor{black}{with and without tasks}. The reported times for macro-particle operations are measured for 1500 iterations and averaged over the MPI processes. }
\label{fig:radiation_pressure2D}
\end{figure}

\textcolor{black}{Figure \ref{fig:radiation_pressure2D} reports the results of a performance study corresponding to the one of Fig. \ref{fig:uniform_plasma2D}, but performed with this second benchmark.  Comparing the two Figures, the effects of a task formulation on a load imbalanced case become evident.}

A first observation, valid also for the first case study, is that the time spent on macro-particle operations is almost constant varying \texttt{\textcolor{black}{bin\_x\_size}} for the version without tasks. 

In this unbalanced case with relatively large patches the version with tasks (\textcolor{blue}{``Tasks ON"}) displays a better scaling than the \textcolor{red}{``Tasks OFF"} version with the considered number of \textcolor{black}{CPU cores}.  For the latter version the time spent on macro-particle operations is almost unchanged varying the number of \textcolor{black}{CPU cores}. This could be expected remembering that the main \texttt{for} loop in Algorithm \ref{ompfor} in that version consists on iterating on a total of 64 patches.  A number of 16, 32, and 64 \textcolor{black}{CPU cores} was considered to distribute these 64 iterations of the for loop, thus practically no scaling is observed passing from 32 to 64 \textcolor{black}{CPU cores}.  Instead, the version with tasks decouples the scheduling of patches and species, i.e. $64$ patches$\times2$ species $=128$ iterations of the double for loop in Algorithm \ref{omptask} are distributed among the threads. Besides, the finer decomposition with bins and the decoupling between thread number and PIC operator offers additional flexibility \textcolor{black}{in} the scheduling of operations at runtime.  

From Fig. \ref{fig:radiation_pressure2D} we also note that the gain in performance of the \textcolor{blue}{``Tasks ON"} version is greater when changing from 8 to 16 OpenMP threads if the number of bins is larger (i.e.  when \texttt{\textcolor{black}{bin\_x\_size}} is smaller): with more bins, more tasks are generated to be executed in parallel by a higher number of OpenMP threads, resulting in a better load balancing and consequently in a better strong scaling.

This additional flexibility in the version with tasks and the lack of scaling for the version without tasks yields for the task version a maximum speed-up of 1.3, 2.4,  4.1 with 16, 32, 64 \textcolor{black}{CPU cores} respectively.

\subsection{Task scheduling visualization}
\begin{figure}[!ht]
\centering
\includegraphics[scale=0.6]{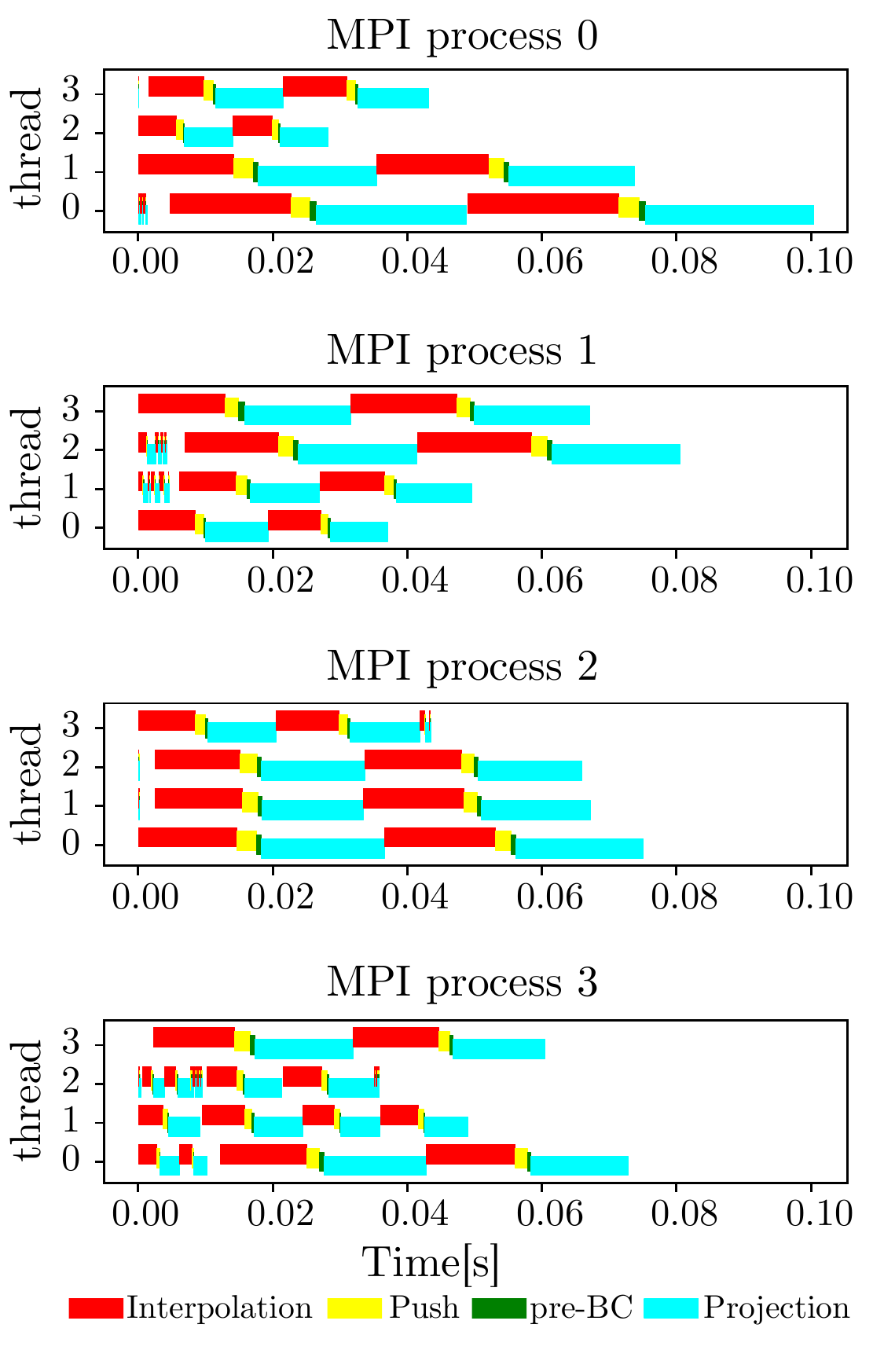}
\caption{Scheduling of macro-particle operations for the 2D radiation pressure benchmark, 4 MPI processes and 4 OpenMP threads, \texttt{\textcolor{black}{bin\_x\_size}} $=64$ (i.e. 1 bin per patch), during iteration 1200, with \textcolor{red}{``Tasks OFF"} version.}
\label{fig:develop_tracing2D}
\end{figure}

\begin{figure}[!ht]
\centering
\includegraphics[scale=0.6]{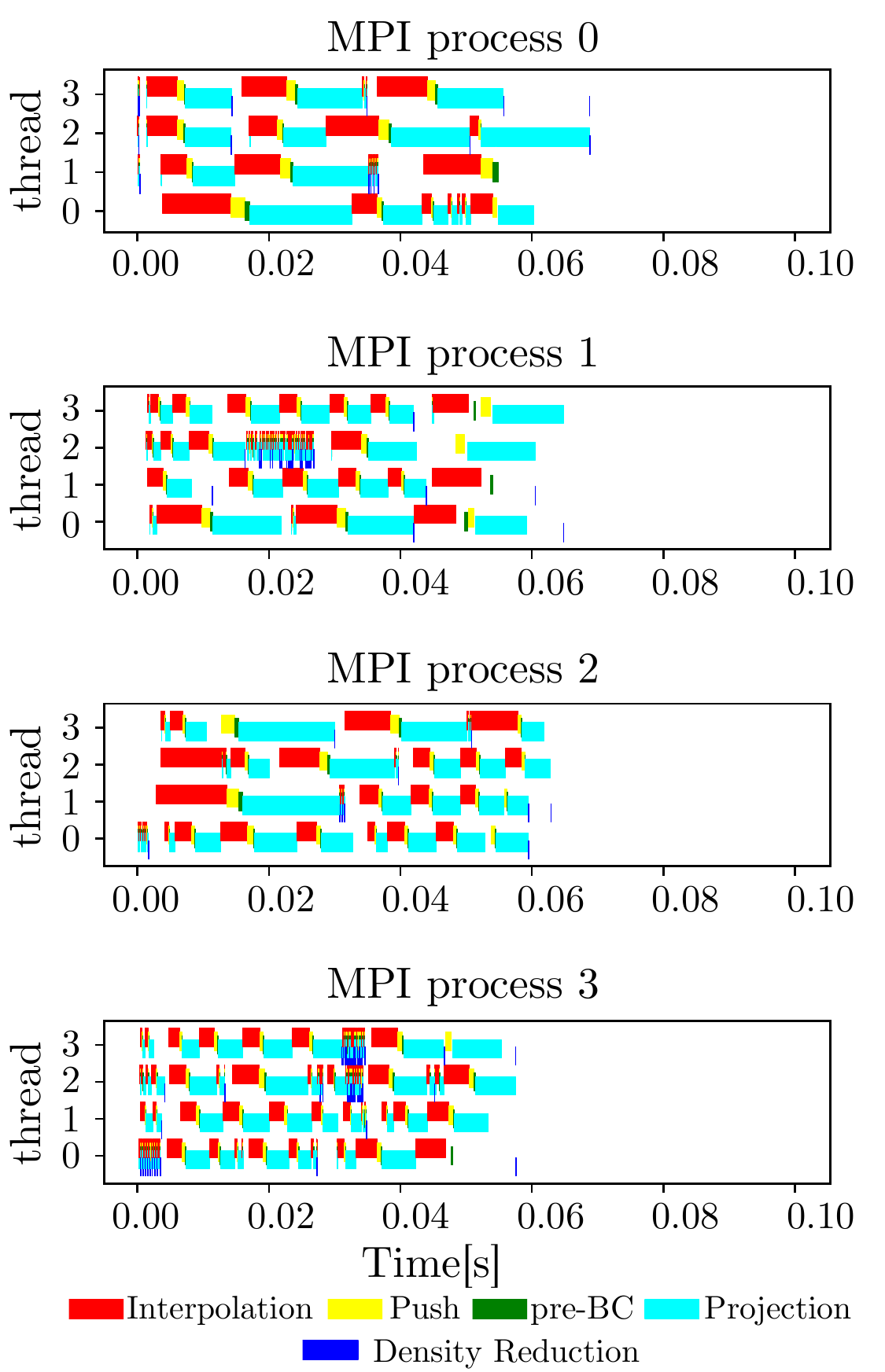}
\caption{Scheduling of macro-particle operations for the 2D radiation pressure benchmark, 4 MPI processes and 4 OpenMP threads, \texttt{\textcolor{black}{bin\_x\_size}} $=16$ (i.e. 4 bins per patch), during iteration 1200, with \textcolor{blue}{``Tasks ON"} version. The horizontal axis has been extended to the same maximum value of the horizontal axis in Fig. \ref{fig:develop_tracing2D} to facilitate the comparison. }
\label{fig:task_tracing2D}
\end{figure}

It is interesting to observe more in detail the task scheduling of macro-operations in the radiation pressure benchmark at a given iteration.  Figures  \ref{fig:develop_tracing2D}, \ref{fig:task_tracing2D} depict such scheduling through an $ad-hoc$ diagnostic.  This case with few large patches was chosen also to make these images more readable. Still, for the sake of clarity the following Figures do not show the \texttt{ipatch}, \texttt{ispec}, \texttt{ibin} for which the displayed operations are executed, but only the type of operation, e.g. Interpolation, Push. In each Figure, the scheduling of macro-particle operations for all the MPI ranks is shown at iteration 1200, for the simulations with 4MPI processes and 4 OpenMP threads, for the versions \textcolor{red}{``Tasks OFF"} and \textcolor{blue}{``Tasks ON"} . A bin size \texttt{\textcolor{black}{bin\_x\_size}} $=16$ has been chosen for the simulation with tasks.  Since for the version without tasks the execution time does not change with \texttt{\textcolor{black}{bin\_x\_size}}, the case \texttt{\textcolor{black}{bin\_x\_size}} $=64$ is reported.

From Fig. \ref{fig:develop_tracing2D} it can be inferred that the coarser grain of the patch decomposition and the sequential execution of the PIC operators does not always allow an efficient scheduling of these operations with the \textcolor{red}{``Tasks OFF"} version.

With the \textcolor{blue}{``Tasks ON"} version, the finer domain decomposition given by the bins and the increased flexibility in the scheduling allows to execute the involved operations more efficiently.

This small-scale example was chosen to illustrate the differences in the scheduling of operations with and without tasks.  The large patches did not allow the dynamic OpenMP scheduling to perform efficiently, so one strategy to improve the performances without tasks is to reduce the size of the patches. In the next section a more realistic example will be shown, where the patch size is near to the minimum size allowed by the interpolation/projection stencils. In that case, the possibility to have \textcolor{black}{bins smaller than the stencils' size} and \textcolor{black}{to have more freedom in the distribution of the operations to the OpenMP threads} is used by the version with tasks to efficiently schedule the operations.

\section{Plasma expansion benchmark}\label{plasma expansion}
In the simplified case of radiation pressure expansion of the previous section it was inferred that reducing the grain of domain decomposition gives an advantage in the scheduling of operations. That simple case study with large patches was used to illustrate the conceptual advantages of tasks in scheduling in a small scale case.  The low number of \textcolor{black}{CPU cores}, patches and species also helped to visualize more clearly the advantage of the task formulation in the scheduling of operations.  

\begin{figure}[H]
\centering
\includegraphics[scale=0.45]{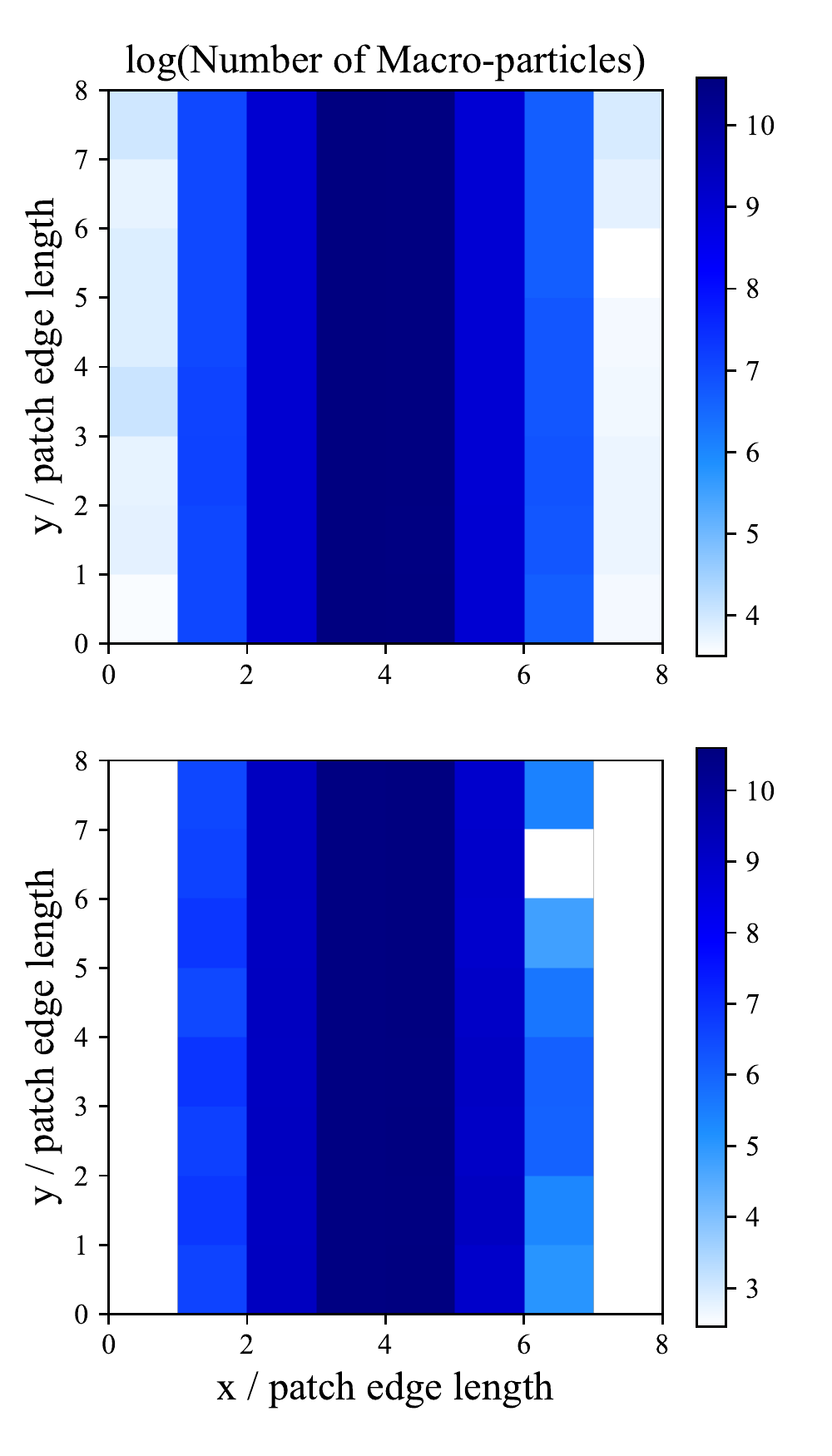}
\caption{Number of macro-particles per patch at half the window's $z$ length in the 3D plasma expansion benchmark at iteration 2000, the end of the simulation. Top panel: electrons; Bottom panel: ions. The $x$ and $y$ axes are scaled by the cubic patches' edge lengths to show the border of the patches. The white in the colormap corresponds to absence of particles} 
\label{fig:plasma_expansion3D_load}
\end{figure}

In this section a performance study with a case that is more representative of the typical use of \Smilei is presented: a 3D plasma expansion case with small patches. This allows the dynamic load balancing at the MPI level to improve the code performances.
In the considered case study a thin preionized plasma slab of thickness $0.2$ $\lambda_r$, where $\lambda_r$ is an arbitrary reference length, is initialized at the center of the $x$ axis of a simulation window with $8\times8\times8$ patches, each with size $8\times8\times8$ cells. Since interpolation of order 2 is used (a stencil with an extent of 3 points),  the code does not allow a patch size smaller than 6 in a given direction (3 points stencil + 3 ghost cells), thus the grain of domain decomposition is already near the allowed limit. The plasma slab's \textcolor{black}{transverse} extent reaches the $y$ and $z$ borders of the simulation window. The plasma is composed of electrons with temperature $0.05$ $m_ec^2$ and cold ions, initially distributed uniformly in the plasma slab with density $100$ $n_c$ and 64 regularly-spaced macro-particles per cell.  The spatial resolution in all directions is set at $\Delta x=\Delta y=\Delta z=0.1$ $c/\omega_r$, while the timestep is set at $\Delta t= 0.9/\sqrt{3}$ $\Delta x/c$.
The reported times spent on macro-particle operations are the total ones measured for $2000$ iterations,  iteration at which the plasma electrons have reached the $x$ borders of the simulation window.
Dynamic load balancing at the MPI level is activated at the start and executed every 100 iterations.
To show the load imbalance of the macro-particle distribution along the patches, Fig.\ref{fig:plasma_expansion3D_load} reports a histogram of the number of macro-particles per patch in a plane at half the $z$ length of the simulation window at the end of the simulation. At $t=0$ the equivalent histogram would display a slab of macro-particles similar to Fig. \ref{fig:radiation_pressure2D_rho} (left panel), but centered at half the window's $x$ length. From the Figure it can be inferred that even if the plasma has an expansion towards the borders of the simulation, the number of macro-particles per patch still differs by orders of magnitude between the central patches and those at the borders.

\begin{figure}[H]
\centering
\includegraphics[scale=0.65]{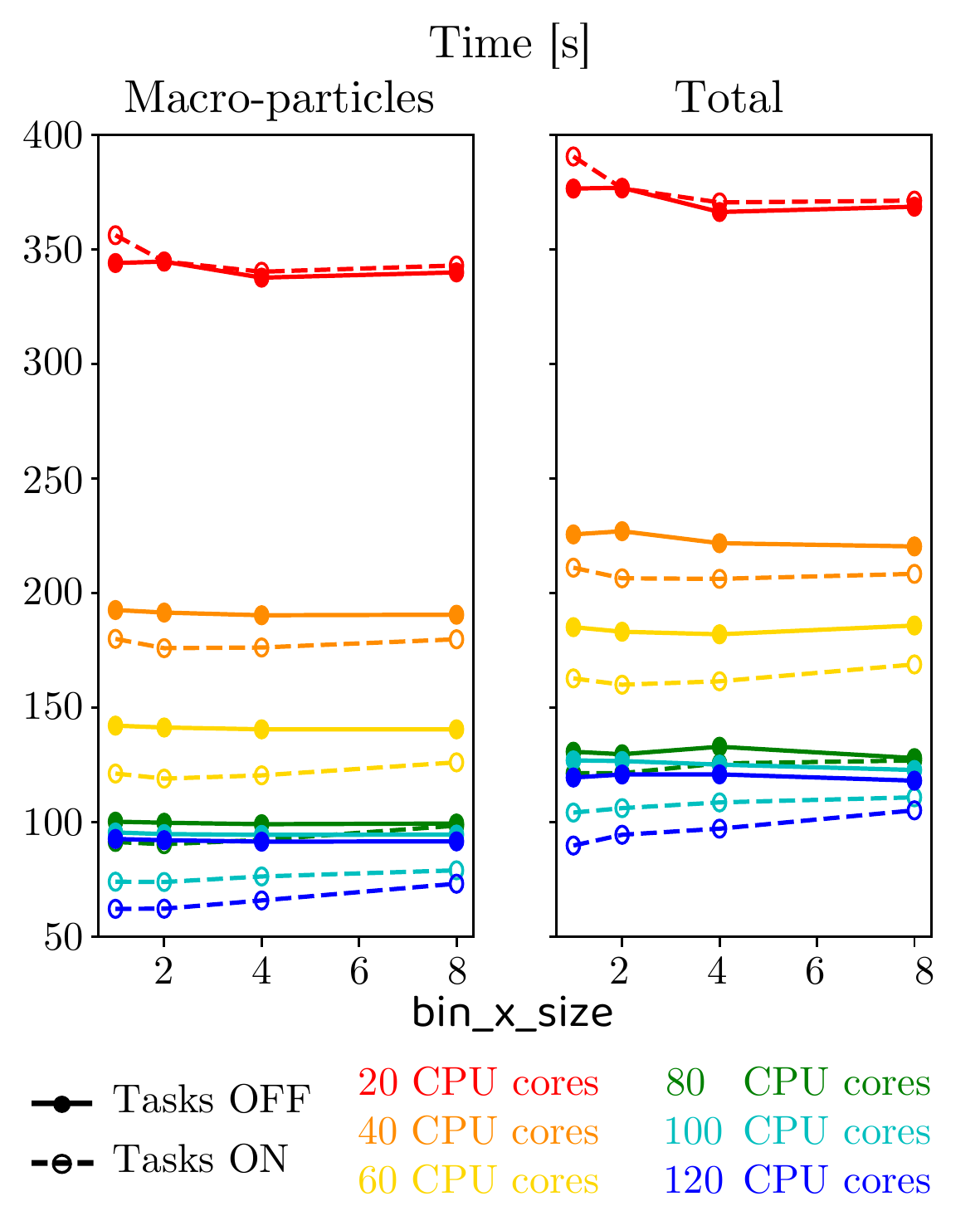}
\caption{Scan for the 3D plasma expansion benchmark. Left panel: time for the macro-particle operations. Right panel: total time for the simulation. The reported times are measured for 2000 iterations; in the case of the time for macro-particle operations they are averaged over the MPI processes.  }
\label{fig:plasma_expansion3D}
\end{figure}

\begin{figure}[H]
\centering
\includegraphics[scale=0.6]{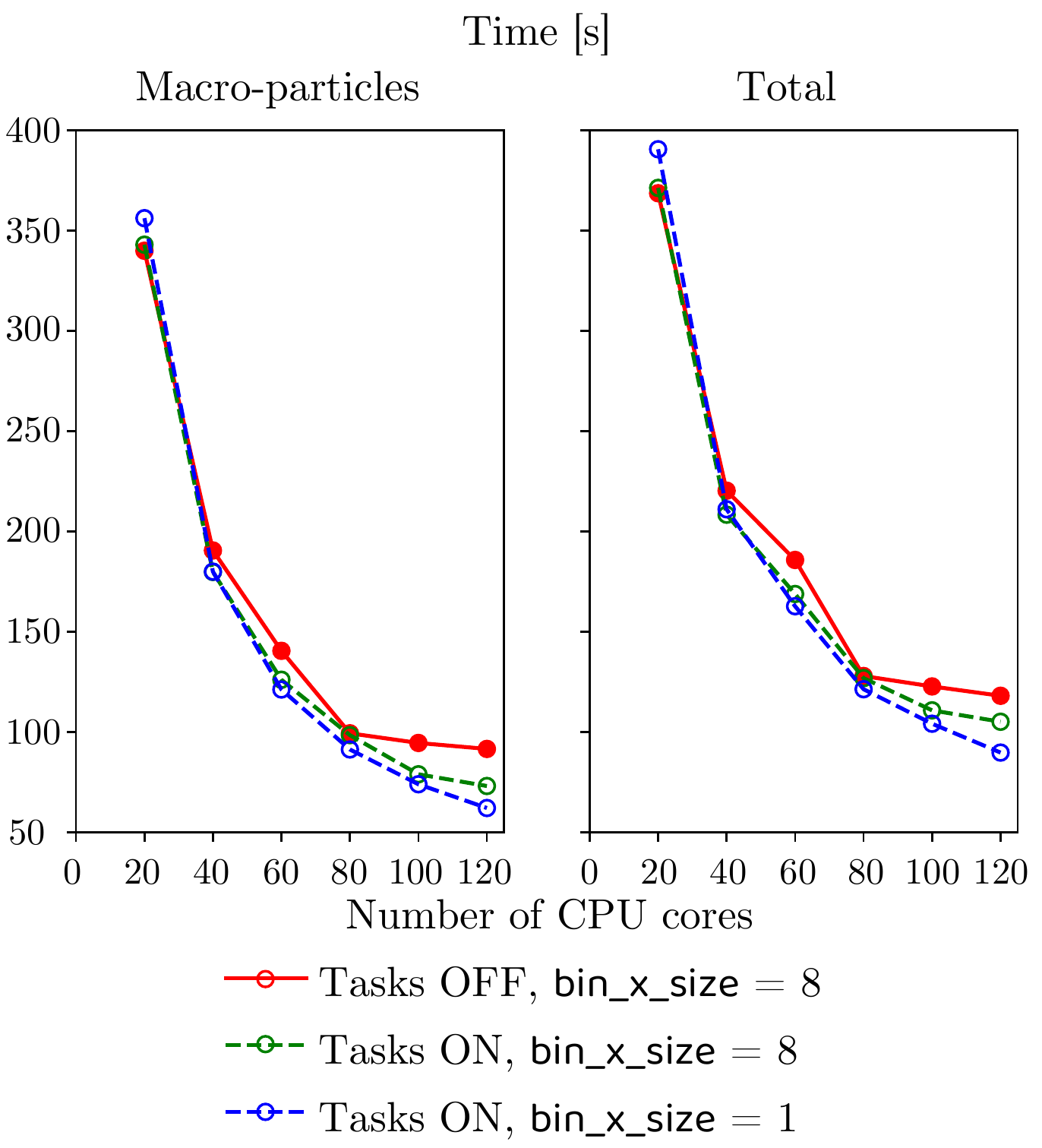}
\caption{Number of cores vs computation time. Left panel: time for the macro-particle operations. Right panel: total time for the simulation. The reported times are measured for 2000 iterations; in the case of the time for macro-particle operations they are averaged over the MPI processes. }
\label{fig:plasma_expansion3D_cpu_scan}
\end{figure}

\begin{figure}[H]
\centering
\includegraphics[scale=0.55]{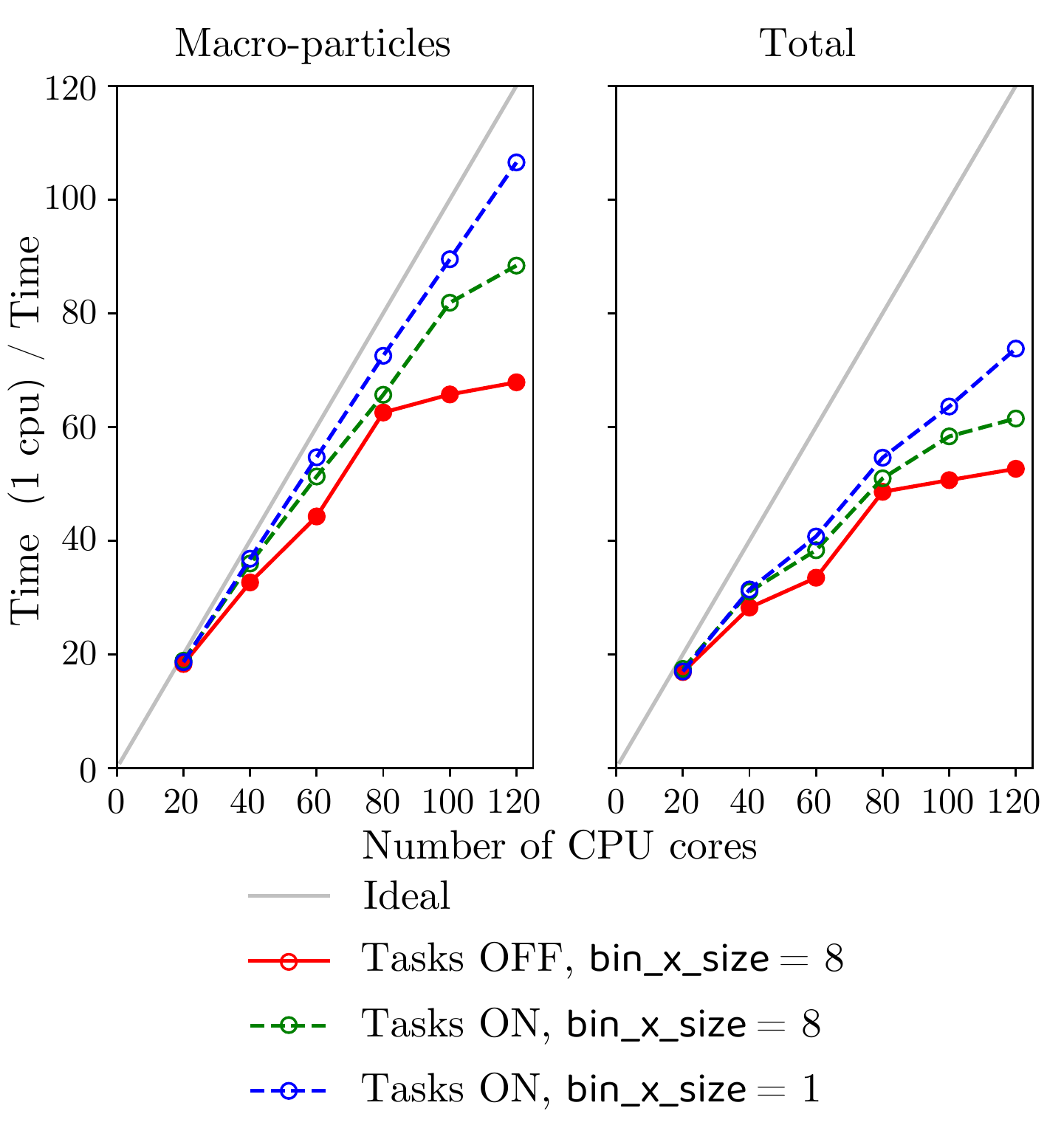}
\caption{Strong scaling, i.e. number of cores vs computation time, divided by the same time on 1 \textcolor{black}{CPU core}, with the same version and \texttt{\textcolor{black}{bin\_x\_size}}. Left panel: time for the macro-particle operations. Right panel: total time for the simulation. The reported times are the total ones measured for 2000 iterations; in the case of the time for macro-particle operations they are averaged over the MPI processes.  
Time for macro-particle operations on 1 \textcolor{black}{CPU core} for 1) \textcolor{blue}{Tasks OFF,  \texttt{\textcolor{black}{bin\_x\_size}} $=8$: $6219$ s}; 2) \textcolor{teal}{Tasks ON,  \texttt{\textcolor{black}{bin\_x\_size}} $=8$: $6469$ s}; 3) \textcolor{red}{Tasks ON  \texttt{\textcolor{black}{bin\_x\_size}} $=1$: $6628$ s}.
Total time on 1 \textcolor{black}{CPU core} for 1) \textcolor{blue}{Tasks OFF,  \texttt{\textcolor{black}{bin\_x\_size}} $=8$: $6494$ s}; 2) \textcolor{teal}{Tasks ON,  \texttt{\textcolor{black}{bin\_x\_size}} $=8$: $6742$ s}; 3) \textcolor{red}{Tasks ON  \texttt{\textcolor{black}{bin\_x\_size}} $=1$: $7014$ s}.
}
\label{fig:plasma_expansion3D_strong_scaling}
\end{figure}

For this study the number of \textcolor{black}{CPU cores} and the bin width along the $x$ direction \texttt{\textcolor{black}{bin\_x\_size}} have been varied in the intervals $[20,40,60,80,100,120]$ and $[1,2,4,8]$ respectively.  For each simulation 20 OpenMP threads were used for each MPI process. Figure \ref{fig:plasma_expansion3D} shows the time spent on macro-particle operations and the total simulation time, with tasks and without tasks, varying the number of \textcolor{black}{CPU cores} and the bin width along the $x$ direction \texttt{\textcolor{black}{bin\_x\_size}}.  Figure \ref{fig:plasma_expansion3D_cpu_scan} shows a subset of the simulations shown in the previous Figure: since the reported times in the version without tasks remain similar varying \texttt{\textcolor{black}{bin\_x\_size}} only the simulations with \texttt{\textcolor{black}{bin\_x\_size}} $=8$ are shown.  For the sake of clarity only the simulations with \texttt{\textcolor{black}{bin\_x\_size}} $=1$, $8$ are reported for the version with tasks. With $20$ \textcolor{black}{CPU cores} the version without tasks is \textcolor{black}{slightly} quicker than the one with tasks, but increasing the number of \textcolor{black}{CPU cores} the tasks start to show their advantage. Additionally, from this Figure it is already apparent that the version without tasks starts to saturate its speed-up after 80 \textcolor{black}{CPU cores}, which is more evident in the strong scaling reported in Fig. \ref{fig:plasma_expansion3D_strong_scaling}. In this Figure the same simulations of Fig. \ref{fig:plasma_expansion3D_cpu_scan} are reported, but the time spent on macro-particle simulations (and total simulation time) are divided by the respective times spent on macro-particle operations ( by the total simulation time) with the same version and \texttt{\textcolor{black}{bin\_x\_size}} on 1 \textcolor{black}{CPU core}. It can be inferred that the tasks improved the strong scaling on this simulation, \textcolor{black}{especially} choosing a finer grain (which would be impossible in the version without tasks at the moment). In the case of \texttt{\textcolor{black}{bin\_x\_size}} $=1$, the observed ratio of the measured time with the task version over the one of the version without tasks is 0.96,1.08,1.17,1.1,1.28,1.49 with 20,40,60,80,100,120 \textcolor{black}{CPU cores} respectively. To understand why the \textcolor{blue}{``Tasks ON"} version has a better strong scaling than \textcolor{red}{``Tasks OFF"}, the number of iterations per \textcolor{black}{CPU core} in their Algorithms should be considered, as mentioned in Section \ref{sec:four_code_versions}. This load-imbalanced benchmark has $8\times8\times8=512$ patches and $2$ species. Thus, with the \textcolor{red}{``Tasks OFF"} version there are 512 iterations in the loop of Algorithm \ref{ompfor} (one for each patch). In the most extreme case there are then on average $512/120\approx 4.3$ iterations per \textcolor{black}{CPU core}. Instead,  with the \textcolor{blue}{``Tasks ON"} version there are at least $512\times2$ iterations defined as tasks for the loop in Algorithm \ref{omptask} to distribute among the \textcolor{black}{CPU cores} (one for each species in each patch). This gives an additional flexibility in the scheduling of macro-particle operations. Besides, reducing the bin size, i.e. increasing the number of bins in each species, the effective number of iterations is multiplied by the number of bins. As can be seen in Fig.\ref{fig:plasma_expansion3D_strong_scaling}, this finer decomposition improves the strong scaling with the task version (\texttt{\textcolor{black}{bin\_x\_size}} $=8$ corresponds to 1 bin, \texttt{\textcolor{black}{bin\_x\_size}} $=1$ corresponds to 8 bins).\\
\indent Figures  \ref{fig:plasma_expansion3D}, \ref{fig:plasma_expansion3D_cpu_scan}, \ref{fig:plasma_expansion3D_strong_scaling} also report the total simulation time in their right panels.  The total simulation time includes the time spent on macro-particle operations (the left panels),  but also e.g. the time spent to solve Maxwell's equations and the time spent on synchronizations \textcolor{black}{and communications.} The latter ones also take into account MPI/OpenMP barriers, the time spent on exchanging border-crossing macro-particles and the values of the fields at the borders of patches, both at the OpenMP thread level and the MPI level. The profiling (not reported here) shows that this ``\textcolor{black}{synchronization/communication} time" constitutes the majority of the time not spent on macro-particle operations. Comparing with the left panels, it can be inferred that its relevance in the total simulation time increases when the number of \textcolor{black}{CPU cores} is increased. The reason is that the time spent on macro-particle operations is reduced with the increased related parallelism and more communications are necessary between the patch/\textcolor{black}{MPI patch collection} borders. Future work may investigate the use \textcolor{black}{of} tasks to overlap the involved communications and the macro-particle operations, or to improve the scheduling of the communications.

\section{Conclusions}
A version of the open source PIC code \Smilei,where the macro-particle operations (Interpolation, Push, \textcolor{black}{preliminary operations for Boundary Conditions,} Projection) are parallelized with OpenMP tasks was described. This task formulation is compatible with the code's pre-existing hybrid MPI+OpenMP parallelization based on patches. A 2D toy benchmark with few large patches was used to show how the tasks introduce an additional flexibility in the scheduling of macro-particle operations and thus improve their scaling in load-imbalanced cases. A visualization of this scheduling with and without tasks was also shown for this simple case. A more realistic load-imbalanced 3D benchmark with many small patches was then reported, highlighting an improved strong scaling. This result was also due to the finer bin decomposition of patches that could be introduced with tasks. This example also showed that future work may focus on the use of tasks to parallelize other operations that become more relevant in the total simulation time when the number of \textcolor{black}{CPU cores} is increased.

\section*{Acknowledgments}
The work presented in this paper has been supported by AIDAS - AI, Data Analytics and Scalable Simulation - which is a Joint Virtual Laboratory gathering the Forschungszentrum Jülich (FZJ) and the French Alternative Energies and Atomic Energy Commission (CEA). AIDAS aims at extending its impact to other relevant institutions in other EU Member States both to HPC-supported application fields and to the development of leading edge HPC technologies in Europe.

The authors also wish to thank the engineers of the cluster Ruche in the Moulon Mesocentre for computer resources and help.

\renewcommand{\baselinestretch}{1.0}

\bibliographystyle{unsrt}
\bibliography{Bibliography}

\end{document}